\documentclass{emulateapj}
\submitted{Accepted for publication in the Astrophysical Journal on 2012 Dec, 15 - Submitted on 2012 Oct, 16}

\def\blender{{\tt BLENDER}}
\def\kepler{\emph{Kepler}}
\def\Kepler{\emph{KEPLER}}
\def\kms{\ifmmode{\rm km\thinspace s^{-1}}\else km\thinspace s$^{-1}$\fi}
\def\ms{\ifmmode{\rm m\thinspace s^{-1}}\else m\thinspace s$^{-1}$\fi}
\def\modif{}

\shortauthors{Fressin et al.}
\shorttitle{\kepler\ FPR and Occurrence of Planets}

\begin{document}

\title{The false positive rate of \Kepler\ and the occurrence of planets}

\author{
Fran\c{c}ois Fressin\altaffilmark{1},
Guillermo Torres\altaffilmark{1},
David Charbonneau\altaffilmark{1},
Stephen T.\ Bryson\altaffilmark{2},
Jessie Christiansen\altaffilmark{2},
Courtney D.\ Dressing\altaffilmark{1}, 
Jon M.\ Jenkins\altaffilmark{2},
Lucianne M.\ Walkowicz\altaffilmark{3} and
Natalie M.\ Batalha\altaffilmark{2}
}+

\altaffiltext{1}{Harvard-Smithsonian Center for Astrophysics,
Cambridge, MA 02138, USA, ffressin@cfa.harvard.edu}
\altaffiltext{2}{NASA Ames Research Center, Moffett Field, CA 94035, USA}
\altaffiltext{3}{Princeton University, Princeton, NJ 08544, USA}

\begin{abstract}

The \kepler\ Mission is uniquely suited to study the frequencies of
extrasolar planets. This goal requires knowledge of the incidence of
false positives such as eclipsing binaries in the background of the
targets, or physically bound to them, which can mimic the photometric
signal of a transiting planet.  We perform numerical simulations of
the \kepler\ targets and of physical companions or stars in the
background to predict the occurrence of astrophysical false positives
detectable by the Mission. Using real noise level estimates, we
compute the number and characteristics of detectable eclipsing pairs
involving main sequence stars and non-main sequence stars or planets,
and we quantify the fraction of those that would pass the \kepler\
candidate vetting procedure. By comparing their distribution with that
of the \kepler\ Objects of Interest (KOIs) detected during the first
six quarters of operation of the spacecraft, we infer the false
positive rate of \kepler\ and study its dependence on spectral type,
candidate planet size, and orbital period.  We find that the global
false positive rate of \kepler\ is 9.4\%, peaking for giant planets
(6--22\,$R_{\earth}$) at 17.7\%, reaching a low of 6.7\% for small
Neptunes (2--4\,$R_{\earth}$), and increasing again for Earth-size
planets (0.8--1.25\,$R_{\earth}$) to 12.3\%.

Most importantly, we also quantify and characterize the distribution
and rate of occurrence of planets down to Earth size with no prior
assumptions on their frequency, by subtracting from the population of
actual \kepler\ candidates our simulated population of astrophysical
false positives. We find that $16.5 \pm 3.6\%$ of main-sequence FGK
stars have at least one planet between 0.8 and 1.25\,$R_{\earth}$ with
orbital periods up to 85 days. {\modif This result is a significant
step towards the determination of eta-earth, the occurrence of
Earth-like planets in the habitable zone of their parent stars.}
There is no significant dependence of the rates of planet occurrence
between 0.8 and 4 Earth radii with spectral type. In the process, we
derive also a prescription for the signal recovery rate of \kepler\
that enables a good match to both the KOI size and orbital period
distribution, as well as their signal-to-noise distribution.


\end{abstract}

\section{Introduction}
\label{sec:introduction}

In February 2011, the \kepler\ Mission produced a catalog of more than
1200 candidate transiting planets, referred to as \kepler\ Objects of
Interest (KOIs). These candidates were identified during the first
four months of operation of the spacecraft \citep{Borucki:11} in its
quest to determine the frequency of Earth-size planets around Sun-like
stars. This unprecedented sample of potential exoplanets has become an
invaluable resource for all manner of statistical investigations of
the properties and distributions of planets around main-sequence
stars.  The most recent \kepler\ release expanded the sample to more
than 2300 candidates \citep{Batalha:12}, based on 16 months of
observation. The analysis of this information must contend, however,
with the fact that not all photometric signals are caused by
planets. Indeed, false positive contamination is typically the main
concern in transit surveys {\modif \citep[see, e.g.,][]{Brown:03}},
including \kepler, because there is a large array of astrophysical
phenomena that can produce small periodic dimmings in the light of a
star that can be virtually indistinguishable from those due to a true
planetary transit. A common example is a background eclipsing binary
falling within the photometric aperture of a \kepler\ target.

Many of the most interesting candidate transiting planets identified
by the \kepler\ Mission cannot be confirmed with current spectroscopic
capabilities, that is, by the detection of the reflex motion of the
star due to the gravitational pull from the planet. Included in this
category are essentially all Earth-size planets, as well as most
super-Earths that are in the habitable zone of Sun-like stars, which
have Doppler signals too small to detect. Faced with this difficulty,
the approach adopted for such objects is statistical in nature and
consists in demonstrating that the likelihood of a planet is much
greater than that of a false positive, a process referred to as
``validation''. The \kepler\ team has made extensive use of a
technique referred to as \blender\ \citep{Torres:04, Torres:11,
Fressin:11} to validate a number of KOIs, including the majority of
the smallest known exoplanets discovered to date.  This technique
requires an accurate knowledge of the target star usually derived from
spectroscopy or asteroseismology, and makes use of other follow-up
observations for the candidate including high spatial resolution
imaging, Spitzer observations when available, and high-resolution
spectroscopy.  The telescope facilities required to gather such
observations are typically scarce, so it is generally not possible to
have these constraints for thousands of KOIs such as those in the
recent list by \cite{Batalha:12}.  For this reason the \kepler\ team
has concentrated the \blender\ efforts only on the most interesting
and challenging cases.

Based on the previous list of \kepler\ candidates by \cite{Borucki:11}
available to them, \cite{Morton:11} (hereafter MJ11) investigated the
false positive rate for KOIs, and its dependence on some of the
candidate-specific properties such as brightness and transit depth.
They concluded that the false positive rate is smaller than 10\% for
most of the KOIs, and less than 5\% for more than half of them. This
conclusion differs significantly from the experience of all previous
transit surveys, where the rates were typically up to an order of
magnitude larger \citep[e.g., $\sim$80\% for the HATNet
survey;][]{Latham:09}. The realization that the \kepler\ rate is much
lower than for the ground-based surveys allowed the community to
proceed with statistical studies based on lists of mere
\emph{candidates}, without too much concern that false positive
contamination might bias the results \citep[e.g.,][]{Howard:12}.

In their analysis MJ11 made a number of simplifying assumptions that
allowed them to provide these first estimates of the false positive
rate for \kepler, but that are possibly not quite realistic enough for
the most interesting smaller signals with lower signal-to-noise ratios
(SNRs). A first motivation for the present work is thus to improve
upon those assumptions and at the same time to approach the problem of
false positive rate determination in a global way, by numerically
simulating the population of blends in greater detail for the entire
sample of \kepler\ targets. A second motivation of this paper is to
use our improved estimates of the false positive rate to extract the
true frequencies of planets of different sizes (down to Earth-size) as
well as their distributions in terms of host star spectral types and
orbital characteristics, a goal to which the \kepler\ Mission is
especially suited to contribute.

Regarding our first objective ---the determination of the false
positive rate of \kepler--- the assumptions by MJ11 we see as
potentially having the greatest impact on their results are the
following:
{\setlength{\leftmargini}{0pt}
\begin{enumerate}
\item The MJ11 false positive rate is based on an assumed 20\% planet
  occurrence, with a power law distribution of planet sizes between
  0.5 and 20\,$R_{\earth}$ (peaking at small planets) independently of
  the orbital period or stellar host characteristics. This is a rather
  critical hypothesis, as the frequency of the smallest or
  longest-period objects in the KOI sample is essentially unknown. The
  adoption of a given planet frequency to infer the false positive
  rate in poorly understood regions of parameter space is risky, and
  may constitute circular reasoning;
\item The scenarios MJ11 included as possible blends feature both
  background eclipsing binaries and eclipsing binaries physically
  associated with the target. However, other configurations that can
  also mimic transit signals were not considered, such as those
  involving larger planets transiting an unseen physical companion or
  a background star. These configurations are more difficult to rule
  out with follow-up observations, and \blender\ studies have shown
  that such scenarios are often the most common blend configuration
  for candidates that have been carefully vetted \citep[see,
  e.g.,][]{Batalha:11, Cochran:11, Ballard:11, Fressin:12b,
  Gautier:12, Borucki:12};
\item A key ingredient in the MJ11 study is the strong constraint
  offered by the analysis of the motion of the centroid of the target
  in and out of transit based the \kepler\ images themselves, which
  can exclude all chance alignments with background eclipsing binaries
  beyond a certain angular separation.  They adopted for this angular
  separation a standard value of 2\arcsec\ based on the single example
  of Kepler-10\,b \citep{Batalha:11}, and then scaled this result for
  other KOIs assuming certain intuitive dependencies with target
  brightness and transit depth. As it turns out, those dependencies
  are not borne out by actual multi-quarter centroid analyses
  performed since, and reported by \cite{Batalha:12};
\item MJ11 did not consider the question of detectability of both
  planets and false positives, and its dependence on the noise level
  for each \kepler\ target star as well as on the period and duration
  of the transit signals;
\item The overall frequency and distribution of eclipsing binaries in
  the \kepler\ field, which enters the calculation of the frequency of
  background blends, has been measured directly by the \kepler\
  Mission itself \citep{Prsa:11, Slawson:11}, and is likely more
  accurate at predicting the occurrence of eclipsing binaries in the
  solar neighborhood than inferences from the survey of non-eclipsing
  binary stars by \cite{Raghavan:10}, used by MJ11.
\end{enumerate}}
The above details have the potential to bias the estimates of the
false positive rate by factors of several, especially for the smaller
candidates with low signal-to-noise ratios. Indeed, recent
observational evidence suggests the false positive rate may be
considerably higher than claimed by MJ11. In one example,
\cite{Demory:11} reported that a significant fraction (14\%) of the
115 hot Jupiter candidates they examined, with radii in the
8--22\,$R_{\earth}$ range, show secondary eclipses inconsistent with
the planetary interpretation. The false positive rate implied by the
MJ11 results for the same radius range is only 4\%.
\cite{Santerne:12} recently conducted an extensive spectroscopic
follow-up campaign using the HARPS and SOPHIE spectrographs, and
observed a number of hot Jupiters with periods under 25 days, transit
depths greater than 0.4\%, and host star magnitudes $K\!p < 14.7$ in
the \kepler\ bandpass. They reported finding a false positive rate of
$34.8 \pm 6.5$\%. In contrast, the MJ11 results imply an average value
of 2.7\% for the same period, depth, and magnitude ranges, which is a
full order of magnitude smaller.

We point out that \cite{Morton:12} recently published an automated
validation procedure for exoplanet transit candidates based on a
similar approach as the MJ11 work, with the improvement that it now
considers background stars transited by planets as an additional
source of potential blends (item \#2 above). However, the
\cite{Morton:12} work does not quantify the global false positive rate
of \kepler, but focuses instead on false alarm probabilities for
individual transit candidates. The methodology makes use of
candidate-specific information such as the transit shape that was not
explicitly considered in the actual vetting procedure used by the
\kepler\ team to generate the KOI list. Here we have chosen to
emulate the \kepler\ procedures to the extent possible so that we
may make a consistent use of the KOI list as published.

Regarding the second goal of our work ---to determine the rate of
occurrence of planets of different sizes--- a main concern that such
statistical studies must necessarily deal with is the issue of
detectability. Only a fraction of the planets with the smallest sizes
and the longest periods have passed the detection threshold of
\kepler.  The studies of \cite{Catanzarite:11} and \cite{Traub:12},
based on the KOI catalog of \cite{Borucki:11}, assumed the sample of
Neptune-size and Jupiter-size planets to be both complete (i.e., that
all transiting planets have been detected by \kepler) and essentially
free of contamination (i.e., with a negligible false positive
rate). They then extrapolated to infer the occurrence rate of
Earth-size planets assuming it follows the occurrence vs.\ size trend
of larger planets.  The more detailed study of \cite{Howard:12}
focused on the sub-sample of KOIs for which \kepler\ is closer to
completeness, considering only planets larger than 2\,$R_{\earth}$ and
with periods shorter than 50 days. They tackled the issue of
detectability by making use of the Combined Differential Photometric
Precision (CDPP) estimates for each KOI to determine the completeness
of the sample. The CDPP is designed to be an estimator of the noise
level of \kepler\ light curves on the time scale of planetary
transits, and is available for each star \kepler\ has observed.
\cite{Howard:12} found a rapid increase in planet occurrence with
decreasing planet size that agrees with the predictions of the
core-accretion formation scenario, but disagrees with population
synthesis models that predict a dearth of objects with super-Earth and
Neptune sizes for close-in orbits. They also reported that the
occurrence of planets between 2 and 4\,$R_{\earth}$ in the \kepler\
field increases linearly with decreasing effective temperature of the
host star. Their results rely to some degree on the conclusions of
MJ11 regarding the rate of false positives of the sub-sample of KOIs
they studied, and it is unclear how the issues enumerated above might
affect them.  {\modif In an independent study \cite{Youdin:11}
developed a method to infer the underlying planetary distribution from
that of the \kepler\ candidates published by \cite{Borucki:11}.  He
investigated the occurrence of planets down to 0.5~$R_{\earth}$, but
without considering false positives, and relied on the simplified
\kepler\ detection efficiency model from \cite{Howard:12}. He reported
a significant difference in the size distributions of shorter ($P <
7$~days) and longer period planets.}

Aside from the question of detectability, we emphasize that the
occurrence rate of Earth-size planets is still effectively unknown.
\kepler\ is the only survey that has produced a list of Earth-size
planet \emph{candidates}, and because of the issues raised above,
strictly speaking we cannot yet rule out that the false positive rate
is much higher for such challenging objects than it is for larger
ones, even as high as 90\%. Indeed, MJ11 predicted false positive
rates with the \emph{assumption} of an overall 20\% planet frequency,
peaking precisely for the smallest planets. The question of the exact
rate of occurrence of small planets has implications for other
conclusions from \kepler. For example, \cite{Lissauer:12} have
recently shown that most KOIs with multiple candidates (`multis') are
bona-fide planets. However, if Earth-size planets are in fact rare,
then most multis involving Earth-size candidates would likely
correspond to \emph{larger} objects transiting (together) the same
unseen star in the photometric aperture of the target, rather than
being systems comprised of true Earth-size planets.

Our two objectives, the determination of the false positive rate for
\kepler\ and the determination of the occurrence rate of planets in
different size ranges, are in fact interdependent, as described below.
We have organized the paper as follows. In Sect.~\ref{sec:approach} we
summarize the general approach we follow. The details of how we
simulate false positives and quantify their frequency are given in
Sect.~\ref{sec:blends}, separately for each type of blend scenario
including background or physically associated stars eclipsed by
another star or by a planet. At the end of this section we provide an
example of the calculation for one of those
scenarios. Sect.~\ref{sec:fprates} is a summary of the false positive
rates for planets of different sizes in the \kepler\ sample as a
whole. In Sect.~\ref{sec:planets} we describe the study of the
\kepler\ detection rate, and how we estimate the frequencies of
planets.  We derive the occurrence of planets in different size ranges
in Sect.~\ref{sec:occurrenceresults}, where we examine also the
dependence of the frequencies on the spectral type (mass) of the host
star and other properties. In Sect.~\ref{sec:duration}, we study the
distribution of the transit durations of our simulated planet
population.  We discuss the assumptions and possible improvements of
our study in Sect.~\ref{sec:discussion}, and conclude by listing our
main results in Sect.~\ref{sec:conclusions}. Finally, an Appendix
describes how we model the detection process of \kepler\ and how we
infer the exclusion limit from the centroid motion analysis for each
individual target.

\section{General approach}
\label{sec:approach}

\subsection{False positive simulation}

The first part of our analysis is the calculation of the false
positive rate for \kepler\ targets.  We consider here all \kepler\
targets that have been observed in at least one out of the first six
quarters of operation of the spacecraft (Q1--Q6)\footnote{A quarter
corresponds to a period of observation of about three months between
90\arcdeg\ spacecraft rolls designed to keep the solar panels properly
illuminated.}, to be consistent with the selection imposed on the
\cite{Batalha:12} list, which comes to a total of 156,453 stars. For
each of these stars, identified by a unique \kepler\ Input Catalog
(KIC) number \citep{Brown:11}, we have available their estimated mass,
surface gravity $\log g$, and brightness in the \kepler\ bandpass
($K\!p$ magnitude) as listed in the KIC\footnote{A number of biases
are known to exist in the properties listed in the KIC, the
implications of which will be discussed later in
Section~\ref{sec:discussion}.}. Additionally we have their CDPP values
on different timescales and for different quarters, provided in the
Mikulski Archive for Space Telescopes (MAST) at STScI.\footnote{{\tt
http://stdatu.stsci.edu/kepler/}} The CDPP allows us to estimate the
detectability of each type of false positive scenario.

We perform Monte Carlo simulations specific to each \kepler\ target to
compute the number of blends of different kinds that we expect. The
calculations are based on realistic assumptions about the properties
of objects acting as blends, informed estimates about the frequencies
of such objects either in the background or physically associated with
the target, the detectability of the signals produced by these blends
for the particular star in question based on its CDPP, and also on the
ability to reject blends based on constraints provided by the centroid
motion analysis or the presence of significant secondary eclipses.
Blended stars beyond a certain angular distance from the target can be
detected in the standard vetting procedures carried out by the
\kepler\ team because they cause changes in the flux-weighted centroid
position, depending on the properties of the signal.  This constraint,
and how we estimate it for each star, is described more fully in the
Appendix.  The inclusion of detectability and the constraint from
centroid motion analysis are included in order to emulate the vetting
process of \kepler\ to a reasonable degree. 

\subsection{Planet occurrence simulation}
\label{sec:classes}

The second part of our analysis seeks to determine the rate of
occurrence of planets of different sizes. For this we compare the
frequencies and distributions of simulated false positives from the
previous step with those of the real sample of KOIs given by
\cite{Batalha:12}. We interpret the differences between these two
distributions as due to true detectable planets. In order to obtain
the actual rate of occurrence of planets it is necessary to correct
for the fact that the KOI list includes only planets with orbital
orientations such that they transit, and for the fact that some
fraction of transiting planets are not detectable by \kepler.  We
determine these corrections by means of Monte-Carlo simulations of
planets orbiting the \kepler\ targets, assuming initial distributions
of planets as a function of planet size and orbital period
\citep{Howard:12}.  We then compare our simulated population of
detectable transiting planets with that of KOIs minus our simulated
false positives population, and we adjust our initial assumptions for
the planetary distributions. We proceed iteratively to obtain the
occurrence of planets that provides the best match to the KOI list.
This analysis permits us to also study the dependence of the
occurrence rates on properties such as the stellar mass.

For the purpose of interpreting both the false positive rates and the
frequencies of planets, we have chosen to separate planets into five
different size (radius $R_p$) ranges:

\begin{itemize}
\item Giant planets: $6\,R_{\earth} < R_p \leq 22\,R_{\earth}$
\item Large Neptunes: $4\,R_{\earth} < R_p \leq 6\,R_{\earth}$
\item Small Neptunes: $2\,R_{\earth} < R_p \leq 4\,R_{\earth}$
\item Super-Earths: $1.25\,R_{\earth} < R_p \leq 2\,R_{\earth}$
\item Earths: $0.8\,R_{\earth} < R_p \leq 1.25\,R_{\earth}$
\end{itemize}

An astrophysical false positive that is not ruled out by centroid
motion analysis and that would produce a signal detectable by \kepler\
with a transit depth corresponding to a planet in any of the above
categories is counted as a viable blend, able to mimic a true planet
in that size range.  These five classes were selected as a compromise
between the number of KOIs in each group and the nomenclature proposed
by the \kepler\ team \citep{Borucki:11,Batalha:12}. In particular, we
have subdivided the 2--6\,$R_{\earth}$ category used by the \kepler\
team into ``small Neptunes'' and ``large Neptunes''. \cite{Howard:12}
chose the same separation in their recent statistical study of the
frequencies of planets larger than 2\,$R_{\earth}$, in which they
claimed that the smallest planets (small Neptunes) are more common
among later-type stars, but did not find the same trend for larger
planets. Our detailed modeling of the planet detectability for each
KOI, along with our new estimates of the false positive rates among
small planets, enables us to extend the study to two smaller classes
of planets, and to investigate whether the claims by \cite{Howard:12}
hold for objects as small as the Earth.

\section{False positives}
\label{sec:blends}

There exists a wide diversity of astrophysical phenomena that can lead
to periodic dimming in the light curve of a \kepler\ target star and
might be interpreted as a signal from a transiting planet.  These
phenomena involve other stars that fall within the photometric
aperture of the target and contribute light. One possibility is a
background star (either main-sequence or giant) eclipsed by a smaller
object; another is a main-sequence star physically associated with the
target and eclipsed by a smaller object. In both cases the eclipsing
body may be either a smaller star or a planet. There are thus four
main categories of false positives, which we consider separately
below.  We discuss also the possibility that the eclipsing objects may
be brown dwarfs or white dwarfs.

We make here the initial assumption that all KOIs that have passed the
nominal detection threshold of \kepler\ are due to astrophysical
causes, as opposed to instrumental causes such as statistical noise or
aliases from transient events in the \kepler\ light curve. The
expected detection rate ($D_r$) based on a matched filter method used
in the \kepler\ pipeline is discussed by \cite{Jenkins:96}.  It is
given by the following formula, as a function of the signal-to-noise
ratio (SNR) of the signal and a certain threshold, assuming that the
(possibly non-white) observational noise is Gaussian:
\begin{equation}
D_r({\rm SNR}) = 0.5 + 0.5 \, {\rm erf} \left[({\rm SNR} - 7.1)/\sqrt{2}\right].
\label{eq:dr}
\end{equation}
In this expression erf is the standard error function, and the adopted
7.1$\sigma$ threshold was chosen so that no more than one false
positive will occur over the course of the Mission due to random
fluctuations \citep[see][]{Jenkins:10}.
The formula corresponds to the cumulative probability density function
of a zero mean, unit variance Gaussian variable, evaluated at a point
representing the distance between the threshold and the SNR. The
construction of the matched filter used in the \kepler\ pipeline is
designed to yield detection statistics drawn from this
distribution. With this in mind, only 50\% of transits with an SNR of
7.1 will be detected. The detection rates are 2.3\%, 15.9\%, 84.1\%,
97.7\%, and 99.9\% for SNRs of 5.1, 6.1, 8.1 9.1, and 10.1.
Below in Sect.~\ref{sec:detectionmodel} we will improve upon this
initial detection model, but we proceed for now with this prescription
for clarity.

Our simulation of the vetting process carried out by the \kepler\ team
includes two tests that the astrophysical scenarios listed above must
pass in order to be considered viable false positives (that could be
considered a planet candidate): the most important is that they must
not produce a detectable shift in the flux centroid. Additionally, for
blends involving eclipsing binaries, they must not lead to secondary
eclipses that are of similar depth as the primary eclipses (within
3$\sigma$).

\subsection{Background eclipsing binaries}
\label{sec:bs}

We begin by simulating the most obvious type of false positive
involving a background (or foreground) eclipsing binary in the
photometric aperture of the target star, whose eclipses are attenuated
by the light of the target and reduced to a shallow transit-like
event. We model the stellar background of each of the 156,453 KIC
targets using the Besan\c{c}on model of the Galaxy
\citep{Robin:03}.\footnote{The accuracy of the stellar densities
(number of stars per square degree) per magnitude bin provided by this
model has been tested against actual star counts at the center of the
\kepler\ field. The results have been found to be consistent within
15\% (R. Gilliland, priv.\ comm.).} We simulate 1 square degree areas
at the center of each of the 21 modules comprising the focal plane of
\kepler\ (each module containing two CCDs), and record all stars (of
any luminosity class) down to 27$^{\rm th}$ magnitude in the $R$ band,
which is sufficiently close to \kepler's $K\!p$ band for our
purposes. This limit of $R = 27$ corresponds to the faintest eclipsing
binary that can mimic a transit by an Earth-size planet, after
dilution by the target. We assign to each \kepler\ target a background
star drawn randomly from the simulated area for the module it is
located on, and we assign also a coefficient $N_{\rm bs}$ to represent
the average number of background stars down to 27$^{\rm th}$ magnitude
in the 1 square degree region around the target.

Only a fraction of these background stars will be eclipsing binaries,
and we take this fraction from results of the \kepler\ Mission itself,
reported in the work of \cite{Slawson:11}. The catalog compiled by
these authors provides not only the overall rate of occurrence of
eclipsing binaries (1.4\%, defined as the number of eclipsing binaries
found by \kepler\ divided by the number of \kepler\ targets), but also
their eclipse depth and period distributions.  In practice we adopt a
somewhat reduced frequency of eclipsing binaries $F_{\rm eb} = 0.79$\%
that excludes contact systems, as these generally cannot mimic a true
planetary transit because the flux varies almost continuously
throughout the orbital cycle. We consider only detached,
semi-detached, and unclassified eclipsing binaries from the
\cite{Slawson:11} catalog. Furthermore, we apply two additional
corrections to this overall frequency that depend on the properties of
the background star. The first factor, $C_{\rm bf}$, adjusts the
eclipsing binary frequency according to spectral type, as earlier-type
stars have been found to have a significantly higher binary frequency
than later-type stars. We interpolate this correction from Figure~12
of \cite{Raghavan:10}. The second factor, $C_{\rm geom}$, accounts for
the increased chance that a larger star in the background would be
eclipsed by a companion of a given period, from geometrical
considerations. This correction factor is unity for a star of
1\,$R_{\sun}$, and increases linearly with radius. We note that the
scaling law we have chosen does not significantly impact the overall
occurrence of eclipsing binaries established by \cite{Slawson:11}, as
the median radius of a \kepler\ target (0.988\,$R_{\sun}$) is very
close to 1\,$R_{\sun}$. The size of each of our simulated background
stars is provided by the Besan\c{c}on model.

Next, we assign a companion star to each of the 156,453 background
stars drawn from the Besan\c{c}on model, with properties (eclipse
depth, orbital period) taken randomly from the \cite{Slawson:11}
catalog. We then check whether the transit-like signal produced by
this companion has the proper depth to mimic a planet with a radius in
the size category under consideration (Sect.~\ref{sec:classes}),
after dilution by the light of the \kepler\ target (which depends on
the known brightness of the target and of the background star).  If it
does not, we reject it as a possible blend.  Next we compute the SNR
of the signal as described in the Appendix, and we determine whether
this false-positive would be detectable by the \kepler\ pipeline as
follows. For each KIC target observed by \kepler\ between Q1 and Q6 we
compute the SNR as explained in the Appendix, and with
Eq.[\ref{eq:dr}] we derive $D_r$. We then draw a random number between
0 and 1 and compare it with $D_r$; if larger, we consider the false
positive signal to be detectable. We also estimate the fraction of the
$N_{\rm bs}$ stars that would be missed in the vetting carried out by
the \kepler\ team because they do not induce a measurable centroid
motion. This contribution, $F_{\rm cent}$, is simply the fraction of
background stars interior to the exclusion limit set by the centroid
analysis. As we describe in more detail in the Appendix, the size of
this exclusion region is itself mainly a function of the SNR.

Based on all of the above, we compute the number $N_{\rm beb}$ of
background (or foreground) eclipsing binaries in the \kepler\ field
that could mimic the signal of a transiting planet in a given size
range as
$$N_{\rm beb} = \sum_{i=1}^{N_{\rm targ}} N_{\rm bs}\, F_{\rm eb}\,
C_{\rm bf}\, C_{\rm geom}\, T_{\rm depth}\, T_{\rm det}\, T_{\rm
sec}\,F_{\rm cent}\,,$$
where $F$ terms represent fractions, $C$ terms represent corrections
to the eclipsing binary frequency, and $T$ terms are either 0 or 1:
\begin{itemize}
\item $N_{\rm targ}$ is the number of \kepler\ targets observed for at
least one quarter during the Q1--Q6 observing interval considered
here;
\item $N_{\rm bs}$ is the average number of background stars down to
magnitude $R = 27$ in a 1 square degree area around the target star;
\item $F_{\rm eb}$ is the frequency of eclipsing binaries, defined as
the fraction of eclipsing binaries (excluding contact systems) found
by \kepler\ divided by the number of \kepler\ targets, and is 0.79\%
for this work;
\item $C_{\rm bf}$ is a correction to the binary occurrence rate
$F_{\rm eb}$ that accounts for the dependence on stellar mass (or
spectral type), following \cite{Raghavan:10};
\item $C_{\rm geom}$ is a correction to the geometric probability of
eclipse that takes into account the dependence on the size of the
background star;
\item $T_{\rm depth}$ is 1 if the signal produced by the background
eclipsing binary has a depth corresponding to a planet in the size
range under consideration, after accounting for dilution by the
\kepler\ target, or 0 otherwise;
\item $T_{\rm det}$ is 1 if the signal produced by the background
eclipsing binary is detectable by the \kepler\ pipeline, or 0 otherwise;
\item $T_{\rm sec}$ is 1 if the background eclipsing binary does not
show a secondary eclipse detectable by the \kepler\ pipeline, or 0
otherwise. We note that we have also set $T_{\rm sec}$ to 1 if the
simulated secondary eclipses have a depth within $3\sigma$ of that of
the primary eclipse, as this could potentially be accepted by the
pipeline as a transiting object, with a period that is half of the
binary period;
\item $F_{cent}$ is the fraction of the $N_{\rm bs}$ background stars
  that would show no significant centroid shifts, i.e., that are
  interior to the angular exclusion region that is estimated for each
  star, as explained in the Appendix.
\end{itemize}

\subsection{Background stars transited by planets}
\label{sec:bp}

A second type of astrophysical scenario that can reproduce the shape
of a small transiting planet consists of a larger planet transiting a
star in the background (or foreground) of the \kepler\ target.  We
proceed in a similar way as for background eclipsing binaries, except
that instead of assigning a stellar companion to each background star,
we assign a random planetary companion drawn from the actual list of
KOIs by \cite{Batalha:12}.  However, there are three factors that
prevent us from using this list as an unbiased sample of transiting
planets: (1) the list is incomplete, as the shallowest and longest
period signals are only detectable among the \kepler\ targets in the
most favorable cases; (2) the list contains an undetermined number of
false positives; and (3) the occurrence of planets of different sizes
may be correlated with the spectral type of the host star
\citep[see][]{Howard:12}.

While it is fairly straightforward to quantify the incompleteness by
modeling the detectability of signals for the individual KOIs (see the
Appendix for a description of the detection model we use), the biases
(2) and (3) are more difficult to estimate. We adopt a bootstrap
approach in which we first determine the false positive rate and the
frequencies for the larger planets among the KOIs, and we then proceed
to study successively smaller planets.  The only false positive
sources for the larger planet candidates involve even larger (that is,
stellar) objects, for which the frequency and distributions of the
periods and eclipse depths are well known \citep[i.e., from the work
of][]{Slawson:11}.  Comparing the population of large-planet KOIs (of
the `giant planet' class previously defined) to the one of simulated
false positives that can mimic their signals enables us to obtain the
false positive correction factor (2). This, in turn, allows us to
investigate whether the frequency of large planets is correlated with
spectral type (see Sect.~\ref{sec:occurrenceresults}), and therefore
to address bias (3) above. As larger planets transiting background
stars are potential false positive sources for smaller planets, once
we have understood the giant planet population we may proceed to
estimate the false positive rates and occurrence of each of the
smaller planet classes in order of decreasing size, applying
corrections for the biases in the same way as described above.

We express the number $N_{\rm btp}$ of background stars in the
\kepler\ field with larger transiting planets able to mimic the signal
of a true transiting planet in a given size class as:
$$N_{\rm btp} = \sum_{i=1}^{N_{\rm targ}} N_{\rm bs}\, F_{\rm tp}\,
C_{\rm tpf}\, C_{\rm geom}\, T_{\rm depth}\, T_{\rm det}\, F_{\rm
cent}\,,$$
where $N_{\rm targ}$, $N_{\rm bs}$, $C_{\rm geom}$, $T_{\rm depth}$,
$T_{\rm det}$, and $F_{\rm cent}$ have similar meanings as before, and
\begin{itemize}
\item $F_{\rm tp}$ is the frequency of transiting planets in the size
class under consideration, computed as the fraction of suitable KOIs
found by \kepler\ divided by the number of non-giant \kepler\ targets
\citep[defined as those with KIC values of $\log g > 3.6$,
following][]{Brown:11}\footnote{We do not consider background giant
stars transited by a planet as a viable blend, as the signal would
likely be undetectable.};
\item $C_{\rm tpf}$ is a correction factor to the transiting planet
frequency $F_{\rm tp}$ that accounts for the three biases mentioned
above: incompleteness, the false positive rate among the KOIs of
\cite{Batalha:12}, and the potential dependence of $F_{\rm tp}$ on the
spectral type of the background star.
\end{itemize}

\subsection{Companion eclipsing binaries}
\label{sec:hts}

An eclipsing binary gravitationally bound to the target star (forming
a hierarchical triple system) may also mimic a transiting planet
signal since its eclipses can be greatly attenuated by the light of
the target. The treatment of this case is somewhat different from the
one of background binaries, though, as the occurrence rate of
intruding stars eclipsed by others is independent of the Galactic
stellar population. To simulate these scenarios we require knowledge
of the frequency of triple systems, which we adopt from the results of
a volume-limited survey of the multiplicity of solar-type stars by
\cite{Raghavan:10}. They found that 8\% of their sample stars are
triples, and another 3\% have higher multiplicity. We therefore adopt
a total frequency of $8\% + 3\% = 11\%$, since additional components
beyond three merely produce extra dilution, which is small in any case
compared to that of the \kepler\ target itself (assumed to be the
brighter star).  Of all possible triple configurations we consider
only those in which the tertiary star eclipses the close secondary,
with the primary star being farther removed (hierarchical
structure). The two other configurations involving a secondary star
eclipsing the primary or the tertiary eclipsing the primary would
typically not result in the target being promoted to a KOI, as the
eclipses would be either very deep or `V'-shaped. Thus we adopt 1/3 of
11\% as the relevant frequency of triple and higher-multiplicity
systems.

We proceed to simulate this type of false positive by assigning to
each \kepler\ target a companion star (``secondary'') with a random
mass ratio $q$ (relative to the target), eccentricity $e$, and orbital
period $P$ drawn from the distributions presented by
\cite{Raghavan:10}, which are uniform in $q$ and $e$ and log-normal in
$P$. We further infer the radius and brightness of the secondary in
the $K\!p$ band from a representative 3~Gyr solar-metallicity
isochrone from the Padova series of \cite{Girardi:00}.  We then assign
to each of these secondaries an eclipsing companion, with properties
(period, eclipse depth) taken from the catalog of \kepler\ eclipsing
binaries by \cite{Slawson:11}, as we did for background eclipsing
binaries in Sect.~\ref{sec:bs}. We additionally assign a mass ratio
and eccentricity to the tertiary from the above distributions.

To determine whether each of these simulated triple configurations
constitutes a viable blend, we perform the following four tests: (1)
We check whether the triple system would be dynamically stable, to
first order, using the condition for stability given by
\cite{Holman:99}; (2) we check whether the resulting depth of the
eclipse, after accounting for the light of the target, corresponds to
the depth of a transiting planet in the size range under
consideration; (3) we determine whether the transit-like signal would
be detectable, with the same signal-to-noise criterion used earlier;
and (4) we check whether the eclipsing binary would be angularly
separated enough from the target to induce a centroid shift. For this
last test we assign a random orbital phase to the secondary in its
orbit around the target, along with a random inclination angle from an
isotropic distribution, a random longitude of periastron from a
uniform distribution, and a random eccentricity drawn from the
distribution reported by \cite{Raghavan:10}. The semi-major axis is a
function of the known masses and the orbital period. The final
ingredient is the distance to the system, which we compute using the
apparent magnitude of the \kepler\ target as listed in the KIC and the
absolute magnitudes of the three stars from the Padova isochrone,
ignoring extinction. If the resulting angular separation is smaller
than the radius of the exclusion region from centroid motion analysis,
we count this as a viable blend.

The number $N_{\rm ceb}$ of companion eclipsing binary configurations
in the \kepler\ field that could mimic the signal of a transiting
planet in each size category is given by:
$$N_{\rm ceb} =\sum_{i=1}^{N_{\rm targ}} F_{eb/triple}\, C_{\rm
geom}\, T_{\rm stab}\, T_{\rm depth}\, T_{\rm det}\, T_{\rm sec}\,
T_{\rm cent}\,,$$
in which $N_{\rm targ}$, $C_{\rm geom}$, $T_{\rm depth}$, $T_{\rm
sec}$ and $T_{\rm det}$ represent the same quantities as in previous
sections, while
\begin{itemize}
\item $F_{\rm eb/triple}$ is the frequency of eclipsing binaries in
triple systems (or systems of higher multiplicity) in which the
secondary star is eclipsed by the tertiary. As indicated earlier, we
adopt a frequency of eclipsing binaries in the solar neighborhood of
$F_{\rm eb}=0.79$\%, from the \kepler\ catalog by \cite{Slawson:11}.
The frequency of stars in binaries and higher-multiplicity systems has
been given by \cite{Raghavan:10} as $F_{\rm bin} = 44$\% (where we use
the notation `bin' to mean `non-single').  Therefore, the chance of a
random pair of stars to be eclipsing is $F_{\rm eb}/F_{\rm bin} =
0.0079/0.44 = 0.018$. As mentioned earlier we will assume here that
one third of the triple star configurations \citep[with a frequency
$F_{\rm triple}$ of 11\%, according to][]{Raghavan:10} have the
secondary and tertiary as the close pair of the hierarchical system.
$F_{\rm eb/triple}$ is then equal to $F_{\rm eb}/F_{\rm bin} \times
1/3 \times F_{\rm triple}= 0.018 \times 1/3 \times 0.11 =
0.00066$. The two other configurations in which the secondary or
tertiary is eclipsing the primary would merely correspond to target
binaries slightly diluted by the light of a companion, and we assume
here that those configurations would not have passed the \kepler\
vetting procedure.
\item $T_{\rm stab}$ is 1 if the triple system is dynamically stable,
and 0 otherwise;
\item $T_{\rm cent}$ is 1 if the triple system does not induce a
significant centroid motion, and 0 otherwise.
\end{itemize}

\subsection{Companion transiting planets}
\label{sec:htp}

Planets transiting a physical stellar companion to a \kepler\ target
can also produce a signal that will mimic that of a smaller planet
around the target itself. One possible point of view regarding these
scenarios is to \emph{not} consider them a false positive, as a planet
is still present in the system, only not of the size anticipated from
the depth of the transit signal.  However, since one of the goals of
the present work is to quantify the rate of occurrence of planets in
specific size categories, a configuration of this kind would lead to
the incorrect classification of the object as belonging to a smaller
planet class, biasing the rates of occurrence. For this reason we
consider these scenarios as a legitimate false positive.

To quantify them we proceed in a similar way as for companion
eclipsing binaries in the preceding section, assigning a bound stellar
companion to each target in accordance with the frequency and known
distributions of binary properties from \cite{Raghavan:10}, and
assigning a transiting planet to this bound companion using the KOI
list by \cite{Batalha:12}. Corrections to the rates of occurrence
$F_{\rm tp}$ are as described in Sect.~\ref{sec:bp}. The total number
of false positives of this kind among the \kepler\ targets is then
$$N_{\rm ctp} = \sum_{i=1}^{N_{\rm targ}} F_{\rm bin}\, F_{\rm tp}\,
C_{\rm geom}\, C_{\rm tpf}\, T_{\rm depth}\, T_{\rm det}\, T_{\rm
cent}\, T_{\rm stab}\,,$$
where $F_{\rm bin}$ is the frequency of non-single stars in the solar
neighborhood \citep[44\%;][]{Raghavan:10} and the remaining symbols
have the same meaning as before.

Adding up the contributions from this section and the preceding three,
the total number of false positives in the \kepler\ field due to
physically-bound or background/foreground stars eclipsed by a smaller
star or by a planet is $N_{\rm beb} + N_{\rm btp} + N_{\rm ceb} +
N_{\rm ctp}$.

\subsection{Eclipsing pairs involving brown dwarfs and white dwarfs}

Because of their small size, brown dwarfs transiting a star can
produce signals that are very similar to those of giant
planets. Therefore, they constitute a potential source of false
positives, not only when directly orbiting the target but also when
eclipsing a star blended with the target (physically associated or
not). However, because of their larger mass and non-negligible
luminosity, other evidence would normally betray the presence of a
brown dwarf, such as ellipsoidal variations in the light curve (for
short periods), secondary eclipses, or even measurable velocity
variations induced on the target star. Previous Doppler searches have
shown that the population of brown dwarf companions to solar-type
stars is significantly smaller than that of true Jupiter-mass
planets. Based on this, we expect the incidence of brown dwarfs as
false positives to be negligible, and we do not consider them here.

A white dwarf transiting a star can easily mimic the signal of a true
Earth-size planet, since their sizes are comparable. Their masses, on
the other hand, are of course very different. Some theoretical
predictions suggest that binary stars consisting of a white dwarf and
a main-sequence star may be as frequent as main-sequence binary stars
\citep[see, e.g.,][]{Farmer:03}. However, evidence so far from the
\kepler\ Mission seems to indicate that these predictions may be off
by orders of magnitude. For example, none of the Earth-size candidates
that have been monitored spectroscopically to determine their radial
velocities have shown any indication that the companion is as massive
as a white dwarf. Also, while transiting white dwarfs with suitable
periods would be expected to produce many easily detectable
gravitational lensing events if their frequency were as high as
predicted by \cite{Farmer:03}, no such magnification events compatible
with lensing have been detected so far in the \kepler\ photometry (J.\
Jenkins, priv.\ comm.).  For these reasons we conclude that the
relative number of eclipsing white dwarfs among the 196 Earth-size
KOIs is negligible in comparison to other contributing sources of
false positives.

\subsection{Example of a false positive calculation}
\label{sec:example}

To illustrate the process of false positive estimation for planets of
Earth size ($0.8\,R_{\earth} < R_p \leq 1.25\,R_{\earth}$), we present
here the calculation for a single case of a false positive involving a
larger planet transiting a star physically bound to the \kepler\
target (Sect.~\ref{sec:htp}), which happens to produce a signal
corresponding to an Earth-size planet.

In this case selected at random, the target star (KIC 3453569) has a
mass of 0.73\,$M_{\sun}$, a brightness of $K\!p = 15.34$ in \kepler\
band, and a 3-hour CDPP of 133.1 ppm. As indicated above, the star has
a 44\% \emph{a priori} chance of being a multiple system based on the
work of \cite{Raghavan:10}. We assign to the companion a mass and an
orbital period drawn randomly from the corresponding distributions
reported by these authors, with values of 0.54\,$M_{\sun}$ and
67.6~yr, corresponding to a semimajor axis of $a = 18.0$~AU.

This companion has an \emph{a priori} chance $F_{\rm tp}$ of having a
transiting planet equal to the number of KOIs (orbiting non-giant
stars) divided by the number of non-giant stars ($F_{\rm tp} =
2,\!302/132,\!756 = 0.017$) observed by \kepler\ in at least one
quarter during the Q1--Q6 interval.  We assign to this companion star
a transiting planet drawn randomly from the \cite{Batalha:12} list of
KOIs, with a radius of 1.81\,$R_{\earth}$ (a super-Earth) and an
orbital period of 13.7~days. The diluted depth of the signal of this
transiting planet results in a 140 ppm dimming of the combined flux of
the two stars during the transit, that could be incorrectly
interpreted as an Earth-size planet ($0.91\,R_{\earth}$) transiting
the target star. We set $T_{\rm depth} = 1$, as this example pertains
to false positives mimicking transits of Earth-size planets.  Because
of the smaller radius of the stellar companion ($R_{\rm comp} =
0.707$\,$R_{\sun}$, determined with the help of our representative
3-Gyr, solar-metallicity isochrone), that star has a correspondingly
smaller chance that the planet we have assigned to orbit it will
actually transit at the given period, compared to the median
$\sim$1\,$R_{\sun}$ \kepler\ target. The corresponding correction
factor is then $C_{\rm geom} = 0.707$.

The correction $C_{\rm tpf}$ to the rate of occurrence of transiting
planets acting as blends is the most complicated to estimate, as it
relies in part on similar calculations carried out sequentially from
the larger categories of planets, starting with the giant planets, as
described earlier. For this example we make use of the results of
those calculations presented in Sect.~\ref{sec:planets}. The $C_{\rm
tpf}$ factor corrects the occurrence rate of planets in the
1.25--2\,$R_{\earth}$ super-Earth category (since this is the relevant
class for the particular blended planet we have simulated) for three
distinct biases in the KOI list of \cite{Batalha:12}, due to
incompleteness, false positives, and the possible correlation of
frequency with spectral type.

For our simulated planet with $R_p = 1.81\,R_{\earth}$ and $P =
13.7$~days, the incompleteness contribution to $C_{\rm tpf}$ was
estimated by computing the SNR of the signals generated by a planet of
this size transiting each individual \kepler\ target, using the CDPP
of each target.  We find that the transit signal could only be
detected around 54.1\% of the targets, from which the incompleteness
boost is simply $1/0.541 = 1.848$. The contribution to $C_{\rm tpf}$
from the false positive bias was taken from Table~\ref{tab:results} of
Sect.~\ref{sec:fprates} below, which summarizes the results from the
bootstrap analysis mentioned earlier. That analysis indicates that the
false positive rate for super-Earths is $8.8\%$, i.e., $91.2\%$ of the
KOIs in the 1.25--2\,$R_{\earth}$ are transited by true
planets. Finally, the third contributor to $C_{\rm tpf}$ addresses the
possibility that small Neptunes from the KOI list acting as blends may
be more common, for example, around stars of later spectral types than
earlier spectral types. Our analysis in Sect.~\ref{sec:super-earths}
in fact suggests that super-Earths have a uniform occurrence as a
function of spectral type, and we account for this absence of
correlation by setting a value of $1.0$ for the spectral-type
dependence correction factor.  Combining the three factors just
described, we arrive at a value of $C_{\rm tpf} = 1.848 \times 0.912
\times 1.0 = 1.685$.

To ascertain whether this particular false positive could have been
detected by \kepler, we compute its signal-to-noise ratio as explained
in the Appendix, in terms of the size of the small Neptune relative to
the physically-bound companion, the dilution factor from the target,
the CDPP of the target, the number of transits observed (from the
duration interval for the particular \kepler\ target divided by the
period of the KOI), and the transit duration as reported by
\cite{Batalha:12}. The result, ${\rm SNR} = 9.44$ is above the
threshold (determined using the detection recovery rate studied in
section~\ref{sec:detectionmodel}), so the signal would be detectable
and we set $T_{\rm det} = 1$.

With the above SNR we may also estimate the angular size of the
exclusion region outside of which the multi-quarter centroid motion
analysis would rule out a blend.  The value we obtain for the present
example with the prescription given in the Appendix is 1\farcs1.  To
establish whether the companion star is inside or outside of this
region, we compute its angular separation from the target as
follows. First we estimate the distance to the system from the
apparent magnitude of the target ($K\!p = 15.34$) and the absolute
magnitudes of the two stars, read off from our representative
isochrone according to their masses of 0.73 and 0.54\,$M_{\sun}$.
Ignoring extinction, we obtain a distance of 730~pc. Next we place the
companion at a random position in its orbit around the target, using
the known semi-major axis of 18.0~AU and the random values of other
relevant orbital elements (eccentricity, inclination angle, longitude
of periastron) as reported above. With this, the angular separation is
0\farcs02. This is about a factor of 50 smaller than the limit from
centroid motion analysis, so we conclude this false positive would not
be detected in this way (i.e., it remains a viable blend). We
therefore set $T_{\rm cent} = 1$.

Finally, the formalism by \cite{Holman:99} indicates that, to first
order, a hierarchical triple system like the one in this example would
be dynamically unstable if the companion star were within 1.06~AU of
the target. As their actual mean separation is much larger ($a = 18.0$
AU), we consider the system to be stable and we set $T_{\rm
stab}=1$.

Given the results above, this case is therefore a viable false
positive that could be interpreted as a transiting Earth-size
planet. The chance that this would happen for the particular \kepler\
target we have selected is given by the product $F_{\rm bin}\times
F_{\rm tp}\times C_{\rm geom}\times C_{\rm tpf}\times T_{\rm depth}
\times T_{\rm det} \times T_{\rm cent} \times T_{\rm stab} = 44\%
\times 0.017 \times 0.707 \times (1.848\times 0.912 \times 1.0) \times
1 \times 1 \times 1 \times 1 = 0.0089$. We performed similar
simulations for each of the other \kepler\ targets observed between Q1
and Q6, and found that larger planets transiting unseen companions to
the targets are the dominant type of blend in this size class, and
should account for a total of 16.3 false positives that might be
confused with Earth-size planets. Doing the same for all other types
of blends that might mimic Earth-size planets leads to an estimate of
24.1 false positives among the \kepler\ targets.

\section{Results for the false positive rate of \Kepler}
\label{sec:fprates}

The output of the simulations described above is summarized in
Table~\ref{tab:results}, broken down by planet class and also by the
type of scenario producing the false positive. We point out here again
that these results are based on a revision of the detection model of
\kepler\ (Sect.~\ref{sec:detectionmodel}) rather than the \emph{a
priori} detection rate described in Section~\ref{sec:blends}.  An
interesting general result is that the dominant source of false
positives for all planet classes involves not eclipsing binaries, but
instead large planets transiting an unseen companion to the \kepler\
target. This type of scenario is the most difficult to rule out in the
vetting process performed by the \kepler\ team.

For giant planets our simulations project a total of 39.5 false
positives among the \kepler\ targets, or 17.7\% of the 223 KOIs that
were actually identified in this size category. This is significantly
higher than the estimate from MJ11, who predicted a less than 5\%
false positive rate for this kind of objects. The relatively high
frequency of false positives we obtain is explained by the inherently
low occurrence of giant planets in comparison to the other
astrophysical configurations that can mimic their signal. Another
estimate of the false positive rate for giant planets was made
recently by \cite{Santerne:12}, from a subsample of KOIs they followed
up spectroscopically. They reported that $34.8 \pm 6.5$\% of close-in
giant planets with periods shorter than 25 days, transit depths
greater than 0.4\%, and brightness $K\!p < 14.7$ show radial-velocity
signals inconsistent with a planetary interpretation, and are thus
false positives. Adopting the same sample restrictions we obtain a
false positive rate of $29.3 \pm 3.1$\%, in good agreement with their
observational result. This value is significantly larger than our
overall figure of 17.7\% for giant planets because of the cut at $P <
25$ days and the fact that the false positive rate increases somewhat
toward shorter periods, according to our simulations (see
Sect.~\ref{sec:giantplanets}).

For large Neptunes we find that the false positive rate decreases
somewhat to 15.9\%. This is due mainly to the lower incidence of
blends from hierarchical triples, which can only mimic the transit
depth of planets orbiting the largest stars in the sample, and to the
relatively higher frequency of planets of this size in comparison to
giant planets.  The false positive rate decreases further for small
Neptunes and super-Earths, and rises again for Earth-size planets. The
overall false positive rate of \kepler\ we find by combining all
categories of planets is 9.4\%.

All of these rates depend quite strongly on how well we have emulated
the vetting process of \kepler.  We may assess this as follows.  We
begin by noting that \emph{before} the vetting process is applied, the
majority of false positives are background eclipsing binaries.  To
estimate their numbers, we tallied all such systems falling within the
photometric aperture of a \kepler\ target and contributing more than
50\% of their flux (755 cases). Of these, 465 pass our secondary
eclipse test (i.e., they present secondary eclipses that are
detectable, and that have a depth differing by more than 3$\sigma$
from the primary eclipse). After applying the centroid test we find
that only 44.7 survive as viable blends. In other words, our simulated
application of the centroid test rules out $465 - 44.7 \approx 420$
eclipsing binary blends. We may compare this with results from the
actual vetting of \kepler\ as reported by \cite{Batalha:12}, who
indicated that 1093 targets from an initial list of 1390 passed the
centroid test and were included as KOIs, and were added to the list of
previously vetted KOIs from \cite{Borucki:11}.  The difference, $1390
- 1093 = 297$, is of the same order of magnitude as our simulated
results ($\sim$420), providing a sanity check on our background
eclipsing binary occurrence rates as well as our implementation of the
vetting process. This exercise also shows that the centroid test is by
far the most effective for weeding out blends. Even ignoring the test
for secondary eclipses, the centroid analysis is able to bring down
the number of blends involving background eclipsing binaries to only
83.3 out of the original 755 in our simulations, representing a
reduction of almost 90\%.

Due to the complex nature of the simulations it is non-trivial to
assign uncertainties to the false positive rates reported in
Table~\ref{tab:results}, and the values listed reflect our best
knowledge of the various sources that may contribute. Many of the
ingredients in our simulations rely on counts based directly on
\kepler\ observations, such as the KOI list and the \kepler\ eclipsing
binary catalog. For those quantities it is reasonable to adopt a
Poissonian distribution for the statistical error ($\epsilon_{\rm
stat}$). We have also attempted to include contributions from inputs
that do not rely directly on \kepler\ observations. One is the
uncertainty in the star counts that we have adopted from the
Besan\c{c}on model of \cite{Robin:03}. A comparison of the simulated
star densities near the center of the \kepler\ field with actual star
counts (R.\ Gilliland, priv.\ comm.) shows agreement within 15\%. We
may therefore use this as an estimate of the error for false positives
involving background stars ($\epsilon_{\rm back}$). As an additional
test we compared the Besan\c{c}on results with those from a different
Galactic structure model.  Using the Trilegal model of
\cite{Girardi:05} we found that the stellar densities from the latter
are approximately 10\% smaller, while the distribution of stars in
terms of brightness and spectral type is similar (background stars
using Trilegal are only 0.47 mag brighter in the $R$ band, and 100\,K
hotter, on average). We adopt the larger difference of 15\% as our
$\epsilon_{\rm back}$ uncertainty.  

We have also considered the additional uncertainty coming from our
modeling of the detection level of the \kepler\ pipeline
($\epsilon_{\rm detec}$). While we initially adopted a nominal
detection threshold corresponding to the expected detection rate following
\cite{Jenkins:10}, in practice we find that this is somewhat
optimistic, and below we describe a revision of that condition that
provides better agreement with the actual performance of \kepler\ as
represented by the published list of KOIs by \cite{Batalha:12}.
Experiments in which we repeated the simulations with several
prescriptions for the detection limit yielded a typical difference in
the results compared to the model we finally adopted (see below) that
may be used as an estimate of $\epsilon_{\rm detec}$. We note that the
difference between the results obtained from all these detection
prescriptions and those that use the \emph{a priori} detection rate is
approximately three times $\epsilon_{\rm detec}$. This suggests that
the \emph{a priori} detection model may be used to
predict reasonable lower limits for the false positive rates, as well
as the planet occurrence rates discussed later.

Based on the above, we take the total error for our false positive
population involving background stars to be $\epsilon_{\rm tot} =
\sqrt{\epsilon_{\rm stat}^2 + \epsilon_{\rm back}^2 + \epsilon_{\rm
detec}^2}$. We adopt a similar expression for false positives
involving stars physically associated with the target, without the
$\epsilon_{\rm back}$ term.

\section{Computing the planet occurrence}
\label{sec:planets}

The KOI list of \cite{Batalha:12} is composed of both true planets and
false positives. The true planet population may be obtained by
subtracting our simulated false positives from the KOIs. However, this
difference corresponds only to planets in the \kepler\ field that both
transit their host star and that are detectable by \kepler.  In order
to model the actual distribution of planets in each size class and as
a function of their orbital period, we must correct for the geometric
transit probabilities and for incompleteness. Our approach, therefore,
is to not only simulate false positives, as described earlier, but to
also simulate in detail the true planet population in such a way that
the sum of the two matches the published catalog of KOIs, after
accounting for the detectability of both planets and false positives.
The planet occurrence rates we will derive correspond strictly to the
average number of planets per star.

For our planet simulations we proceed as follows. We assign a random
planet to each \kepler\ target that has been observed between Q1 and
Q6, taking the planet occurrences per period bin and size class to be
adjustable variables. We have elected to use the same logarithmic
period bins as adopted by \cite{Howard:12}, to ease comparisons, with
additional bins for longer periods than they considered (up to
$\sim$400 days).  We seek to determine the occurrence of planets in
each of our five planet classes and for each of 11 period bins, which
comes to 55 free parameters. We use the rates of occurrence found by
\cite{Howard:12} for our initial guess (prior), with extrapolated
values for the planet sizes (below 2\,$R_{\earth}$) and periods
(longer than 50 days) that they did not consider in their study. Each
star is initially assigned a global chance of hosting a planet equal
to the sum of these 55 occurrence rates. Our baseline assumption is
that planet occurrence is independent of the spectral type of the host
star, but we later investigate whether this hypothesis is consistent
with the observations (i.e., with the actual distribution of KOI
spectral types; Sect.~\ref{sec:occurrenceresults}).

We have also assumed that the planet sizes are logarithmically
distributed between the size boundaries of each planet class, and that
their periods are distributed logarithmically within each period bin.
For each of our simulated planets we compute the geometric transit
probability (which depends on the stellar radius) as well as the SNR
of its combined transit signals (see the Appendix for details on the
computation of the SNR). We assigned to each planet a random
inclination angle, and discarded cases that are not transiting or that
would not be detectable by \kepler.  We assigned also a random
eccentricity and longitude of periastron, with eccentricities drawn
from a Rayleigh distribution, following \cite{Moorhead:11}. These
authors found that such a distribution with a mean eccentricity in the
range 0.1--0.25 provides a satisfactory representation of the
distribution of transit durations for KOIs cooler than 5100\,K. We
chose to adopt an intermediate value for the mean of the Rayleigh
distribution of $e = 0.175$, and used it for stars of all spectral
types.  Allowing for eccentric orbits alters the geometric probability
of a transit as well as their duration (and thus their SNR).  Finally,
we compare our simulated population of detectable transiting planets
with that of the KOIs minus our simulated false positives population,
and we correct our initial assumption for the distribution of planets
as a function of size and period.

To estimate uncertainties for our simulated true planet population we
adopt a similar prescription as for the false positive rates, and
compute $\epsilon_{\rm tot} = \sqrt{\epsilon_{\rm stat}^2 +
\epsilon_{\rm detec}^2}$, where the two contributions have the same
meaning as before. While the two terms in $\epsilon_{\rm tot}$ have
roughly the same average impact on the global uncertainty, the
statistical error tends to dominate when the number of KOIs in a
specific size and period bin is small, and the detection error is more
important for smaller planets and longer periods.

Modeling the detection limits of \kepler\ is a central component of
the process, as the incompleteness corrections can be fairly large in
some regimes (i.e., for small signals and/or long periods). Exactly
how this is done affects both the false positive rates and the planet
occurrence rates.  We have therefore gone to some effort to
investigate the accuracy of the nominal detection model
(Sect.~\ref{sec:blends}) according to which 50\% of signals are
considered detected by the \kepler\ pipeline if their SNR exceeds a
threshold of 7.1, and 99.9\% of signals are detected for a SNR over
10.1. We describe this in the following.

\subsection{Detection model}
\label{sec:detectionmodel}

\cite{Burke:12} have reported that a significant fraction of the
\kepler\ targets have not actually been searched for transit signals
down to the official SNR threshold of 7.1. This is due to the fact
that spurious detections with SNR over 7.1 can mask real planet
signals with lower SNR in the same light curve.  {\modif
\cite{Pont:06} have shown that time-correlated noise features in
photometric time-series can produce spurious detections well over this
threshold, and that this has significant implications for the yield of
transit surveys.}  We note also that the vetting procedure that led to
the published KOI list of \cite{Batalha:12} involves human
intervention at various stages, and is likely to have missed some
low-SNR candidates for a variety of reasons.  Therefore, the
assumption that most signals with ${\rm SNR} > 7.1$ have been detected
and are present in the KOI list is probably optimistic.  Nevertheless,
this hypothesis is useful in that it sets lower limits for the planet
occurrence rates, and an upper limit for the false positive rates. For
example, the overall false positive rate (all planet classes) we
obtain when following the \emph{a priori} detection rate is 14.9\%,
compared to our lower rate of 9.4\% when using a more accurate model
given below.

\begin{figure}
\vskip 10pt
\begin{center}
\epsscale{1.15} 
\plotone{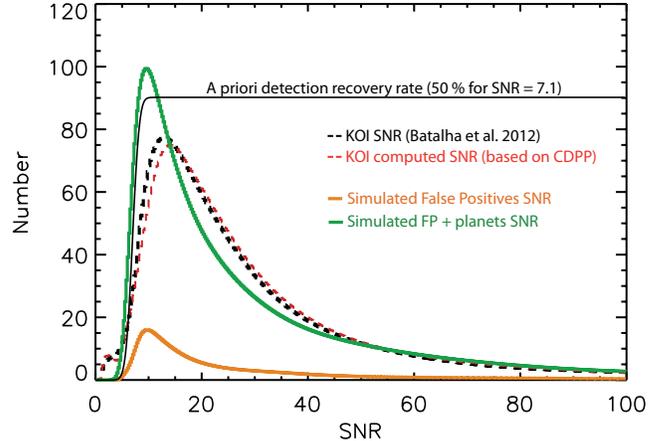}
\vskip 1pt
 \caption{Comparison of signal-to-noise ratio distributions to
 evaluate the detection model (signal recovery rate) of \kepler. The
 dotted black curve corresponds to the KOIs from \cite{Batalha:12},
 with the SNRs from their Table~9 adjusted to correspond to Q1--Q6,
 for consistency with our simulations (see text). There is good
 agreement with the distribution of SNRs computed from the CDPP
 \citep{Christiansen:12} (red curve).  The green curve is the SNR
 distribution of our simulated population of false positives and true
 planets (which by construction matches the size/period distribution
 of actual KOIs). It assumes the nominal \kepler\ detection model in
 which 50\% of the signals with ${\rm SNR} > 7.1$ and 99.9\% of the
 signals with ${\rm SNR} > 10.1$ are detectable (solid black lines).
 The simulated false positives are shown separately by the orange
 curve.  The dotted black curve peaks at a larger SNR than the green
 curve, indicating that the nominal detection model is not accurate:
 the effective recovery threshold must be significantly higher than
 7.1.\label{fig:no_ramp}}
\end{center}
\end{figure}

The clearest evidence that the Kepler team has missed a significant
fraction of the low SNR candidates is seen in
Figure~\ref{fig:no_ramp}. This figure displays the distribution of the
SNRs of the actual KOIs, and compares it with the SNRs for our
simulated population of false positives and true planets. To provide
for smoother distributions we have convolved the individual SNRs with
a Gaussian with a width corresponding to 20\% of each SNR (a kernel
density estimation technique). The SNRs for the KOIs presented by
\cite{Batalha:12} were originally computed based on observations from
Q1 to Q8. For consistency with our simulations, which only use Q1--Q6,
we have therefore degraded the published SNRs accordingly. Also shown
in the figure is the SNR distribution for the KOIs computed with the
prescription described in the Appendix based on the CDPP
\citep{Christiansen:12}. Several conclusions can be drawn.  One is
that there is very good agreement between the SNRs computed by us from
the CDPP (red line) and those presented by \cite{Batalha:12} (adjusted
to Q1--Q6; black line). Importantly, this validates the CDPP-based
procedure used in this paper to determine the detectability of a
signal. It also suggests that the KOI distribution contains useful
information on the actual signal recovery rate of \kepler.  Secondly,
we note that a small number of KOIs ($\sim$70) have SNRs (either
computed from the CDPP or adjusted to Q1--Q6) that are actually below
the nominal threshold of 7.1. Most of these lower SNRs are values we
degraded from Q1--Q8 to Q1--Q6, and others are for KOIs that were not
originally found by the \kepler\ pipeline, but were instead identified
later by further examination of systems already containing one or more
candidates. Thirdly and most importantly, the peak of the SNR
distribution from our simulations (green line in the figure), which by
construction match the size and period distributions of the KOIs and
use the \emph{a priori} detection model, is shifted to smaller values
than the one for KOIs. This suggests that the \emph{a priori}
detection model described in Sect.~\ref{sec:blends} in which 50\% of
the signals with ${\rm SNR} > 7.1$ (and 99.9\% of the signals ${\rm
SNR} > 10.1$) have been detected as KOIs is not quite accurate, and
indicates the detection model requires modification.

\begin{figure}
\vskip 10pt
\begin{center}
\epsscale{1.15} 
\plotone{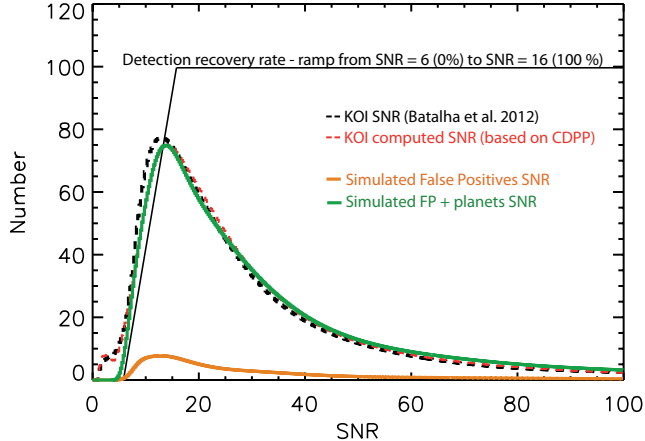}
\vskip 1pt
 \caption{Same as Figure~\ref{fig:no_ramp} but with a detection model
 (solid black lines) ranging linearly from 0\% for candidates with a
 SNR of 6 or lower to 100\% for candidates above a SNR of 16. With
 this model our simulations can match not only the size and period
 distribution of KOIs, but also the distribution of their SNRs.
 \label{fig:ramp}}
\end{center}
\end{figure}

It is not possible to adjust the SNR-dependent recovery rate
simultaneously with the occurrence rates of planets in our
simulations, as the two are highly degenerate. However, we find that
we are able to reach convergence if we assume the following:
\begin{itemize}
  \item the recovery rate is represented by a monotonically rising
	function, rather than a fixed threshold, increasing from zero
	at some low SNR to 100\% for some higher SNR;
  \item the model for the recovery rate function should allow for a
  good match of the distribution of SNRs separately for each of our
  five planet classes.
\end{itemize}
A number of simple models were tried, and we used the Bayesian
Information Criterion (BIC) to compare them and make a choice: ${\rm
BIC} = \chi^2 + k \ln n$. In this expression $\chi^2$ was computed in
the usual way by comparing our simulated SNRs for the population of
false positives and true planets with the one for the KOIs; $k$ is the
number of free parameters of the model, and $n$ is the number of bins
in the histogram of the SNR distributions.  We considered the five
planet categories at the same time and computed the BIC by summing up
the corresponding $\chi^2$ values, with $n$ therefore being the sum of
the number of bins for all classes. The model that provides the best
BIC involves a simple linear ramp for the recovery rate between SNRs
of 6 and 16, in other words, no transit signal with a SNR below 6 is
recovered, and every transit signal is recovered over
16. Figure~\ref{fig:ramp} (to be compared with
Figure~\ref{fig:no_ramp}) shows the much better agreement between the
SNR distribution of our simulated population of false positives and
planets and the SNRs for the KOIs. We adopt this detection model for
the remainder of the paper.

\section{Planet occurrence results}
\label{sec:occurrenceresults}

This section presents the results of our joint simulation of false
positives and true planets for each of the planet categories, and
compares their distributions with those of actual KOIs from
\cite{Batalha:12}.  In particular, since we have assumed no
correlation between the planet frequencies and the spectral type of
the host star (the simplest model), here we examine whether there is
any such dependence, using the stellar mass as a proxy for spectral
type. As in some of the previous figures, we display generalized
histograms computed with a kernel density estimator approach in which
the stellar masses are convolved with a Gaussian function. We adopt a
Gaussian width ($\sigma$) of 20\%, which we consider to be a realistic
estimate of the mass uncertainties in the KIC.

The total frequencies (average number of planets per star) are
reported in Table~\ref{tab:occurrence_period} and
Table~\ref{tab:occurrence_period_cum}. The first table presents the
occurrence of planets of different classes per period bin of equal
size on a logarithmic scale.  The bin with the longest period for
which we can provide an occurrence estimate differs for the different
planet classes. We do not state results for period bins in which the
number of KOIs is less than 1$\sigma$ larger than the number of false
positives; for these size/period ranges the current list of \kepler\
candidates is not large enough to provide reliable values.  Cumulative
rates of occurrence of planets as a function of period are presented
in Table~\ref{tab:occurrence_period_cum}.  Another interesting way to
present planet occurrence results is to provide the number of stars
that have at least one planet in various period ranges.  This requires
a different treatment of KOIs with multiple planet candidates for the
same star.  To compute these numbers, shown in
Table~\ref{tab:stars_with_planets}, we repeated our simulations by
removing from the KOI list the planet candidates beyond the inner one
in the considered size range, for KOIs with multiple candidates.

\subsection{Giant planets (6--22\,$R_{\earth}$)}
\label{sec:giantplanets}

Planet occurrence rates and false positive rates are interdependent.
As described earlier, the bootstrap approach we have adopted to
determine those properties for the different planet classes begins
with the giant planets, as only larger objects (stars) with well
understood properties can mimic their signals. The process then
continues with smaller planets in a sequential fashion.

\begin{figure}
\vskip 10pt
\begin{center}
\epsscale{1.15} 
\plotone{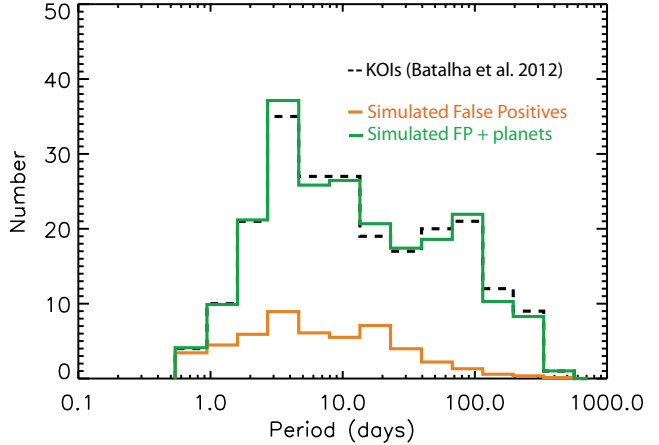}
\vskip 1pt
 \caption{Histogram of the periods of giant planet KOIs from
 \cite{Batalha:12} (dotted black lines), simulated false positives
 (orange), and the sum of simulated false positives and simulated
 planets (green).
 \label{fig:period_6}}
\end{center}
\end{figure}

The frequencies of giant planets per period bin were adjusted until
their distribution added to that of false positives reproduces the
period spectrum of actual KOIs.  This is illustrated in
Figure~\ref{fig:period_6}, where the simulated and actual period
distributions (green and dotted black lines) match on average, though
not in detail because of the statistical nature of the
simulations. The frequency of false positives (orange line) is seen to
increase somewhat toward shorter periods, peaking at $P \sim$ 3--4
days. Similar adjustments to the frequencies by period bin have been
performed successively for large Neptunes, small Neptunes,
super-Earths, and Earth-size planets.

\begin{figure}
\vskip 10pt
\begin{center}
\epsscale{1.15} 
\plotone{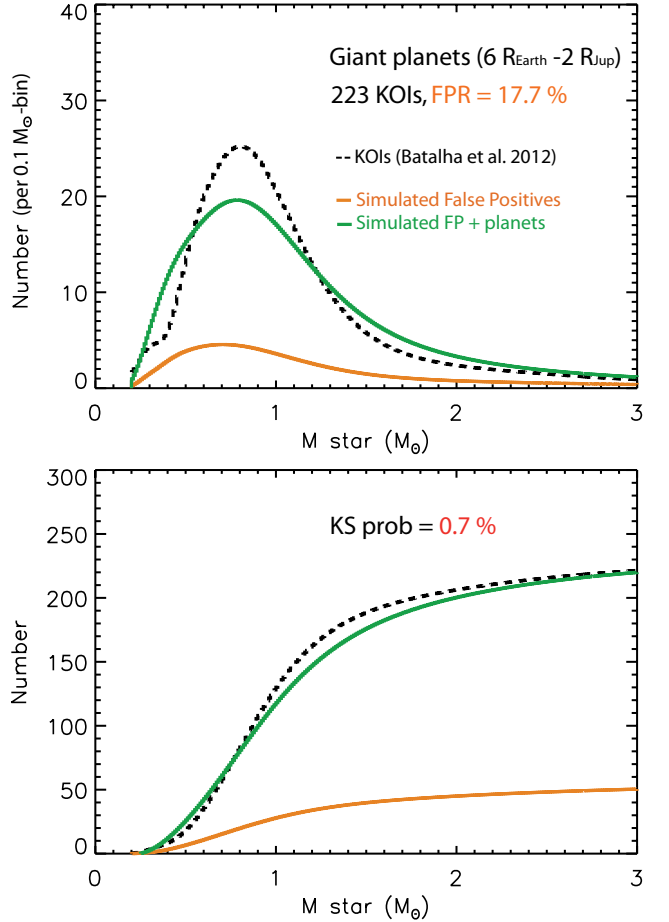}
\vskip 1pt
 \caption{(Top): Generalized histogram of the stellar masses of KOI
 host stars in the giant planet category (dotted black curve),
 compared against the results of our Monte-Carlo simulation of the
 \kepler\ targets observed during Q1--Q6. The total number of KOIs
 with main-sequence host stars in this category (223) and the overall
 false positive rate (${\rm FPR} = 17.7$\%; Table~\ref{tab:results})
 are indicated.  Simulated false positives are shown in orange, and
 the sum of these and the simulated planet population are represented
 in green. The green and dotted black curves are statistically
 different. (Bottom): cumulative distribution of stellar masses from
 the top panel. Our assumed uniform occurrence of giant planets as a
 function of the host star spectral type (mass) results in a
 distribution that is statistically different from the actual KOI
 distribution (K-S probability of $0.7\%$). The KOI distribution
 indicates that G-K stars are more likely to host giant planets than
 F- and M-type stars.\label{fig:mstar_6}}
\end{center}
\end{figure}

We find that the overall frequency of giant planets (planets per star)
in orbits with periods up to 418 days is close to
5.2\%. Figure~\ref{fig:mstar_6} (top panel) displays the distribution
of simulated false positives for this class of planets as a function
of the mass of the host star (orange curve). After adding to these the
simulated planets, we obtain the green curve that represents stars
that either have a true transiting giant planet or that constitute a
false positive mimicking a planet in this category.  The comparison
with the actual distribution of KOIs (black dotted curve) shows
significant differences, with a Kolmogorov-Smirnov (K-S) probability
of 0.7\%. This suggests a possible correlation between the occurrence
of giant planets and spectral type (mass), whereas our simulations
have assumed none.  In particular, the simulations produce an excess
of giant planets around late-type stars, implying that in reality
there may be a deficit for M dwarfs. The opposite seems to be true for
G and K stars. Doppler surveys have also shown a dependence of the
rates of occurrence of giant planets with spectral type. For example,
\cite{Johnson:10} reported a roughly linear increase as a function of
stellar mass, with an estimate of about 3\% for M dwarfs.  We find a
similar figure of $3.6 \pm 1.7$\%. As is the case for the RV surveys,
our frequency increases for G and K stars to $6.1 \pm 0.9$\%, but then
reverses for F (and warmer) stars to $4.3 \pm 1.0$\%, while the
Doppler results suggest a higher frequency approaching 10\%. We point
out, however, that there are rather important differences between the
two samples: (1) the estimates from radial velocity surveys extend out
to orbital separations of 2.5 AU, while our study is only reasonably
complete to periods of $\sim$400 days (1.06 AU for a solar-type star);
(2) the samples are based on two very different characteristics: the
planet mass for RV surveys, and planet radius for \kepler; and (3)
there may well be a significant correlation between the period
distribution of planets and their host star spectral type. Indeed, the
results from the study of planets orbiting A-type stars by
\cite{Bowler:10} provides an explanation for the apparent discrepancy
between RV and \kepler\ results for the occurrence of giant planets
orbiting hot stars: Figure~1 of the above paper shows that no Doppler
planets have been discovered with semimajor axes under 0.6 AU for
stellar masses over 1.5,$M_{\sun}$, creating a `planet desert' in that
region. Thus, the Doppler surveys find the more common longer-period
planets around the hotter stars, and \kepler\ has found the rarer
planets close-in.

Averaged over all spectral types, the frequency of giant planets up to
orbital periods of 418 days is approximately 5.2\%
(Table~\ref{tab:occurrence_period_cum}).

\subsection{Large Neptunes (4--6\,$R_{\earth}$)}
\label{sec:largeneptunes}

\begin{figure}
\vskip 10pt
\begin{center}
\epsscale{1.15} 
\plotone{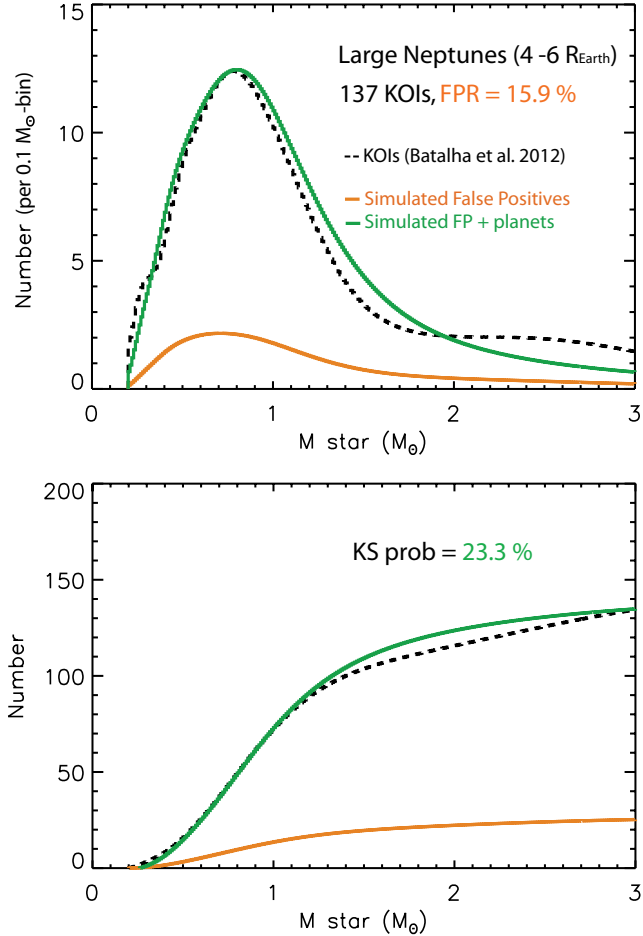}
\vskip 1pt
 \caption{Similar to Fig.~\ref{fig:mstar_6}, for the large
 Neptunes. In this case the mass distributions of KOIs and simulated
 false positives and planets are statistically indistinguishable (K-S
 probability = 23.3\%), supporting the lack of any correlation of the
 planet frequency with spectral type for this class of
 objects.\label{fig:mstar_4}}
\end{center}
\end{figure}

Our simulations result in an overall frequency of large Neptunes with
periods up to 418 days of approximately 3.2\%.  The distribution in
terms of host star mass is shown in Figure~\ref{fig:mstar_4}, and
indicates a good match to the distribution of the large Neptune-size
KOIs from \cite{Batalha:12} (K-S probability of 23.3\%). We
conclude that there is no significant dependence of the occurrence
rate of planets in this class with the spectral type of the host star.

\subsection{Small Neptunes (2--4\,$R_{\earth}$)}
\label{sec:smallneptunes}

The overall rate of occurrence (planets per star) of small Neptunes
rises significantly compared to that of larger planets, reaching 31\%
out to periods of 245 days.  Due to small-number statistics we are
unable to provide reliable estimates for periods as long as those
considered for the larger planets (up to 418 days).  The logarithmic
distribution of sizes we have assumed within each planet category
allows for a satisfactory fit to the actual KOI distributions in each
class (with separate K-S probabilities above 5\%), with the exception
of the small Neptunes. As noted earlier, the increase in planet
occurrence toward smaller radii for these objects is very steep
(Figure~\ref{fig:sizes}). We find that dividing the small Neptunes
into two subclasses (two radius bins of the same logarithmic size:
2--2.8\,$R_{\earth}$, and 2.8--4\,$R_{\earth}$) we are able to obtain
a much closer match to the KOI population (K-S probability of 6\%)
with similar logarithmic distributions within each sub-bin as assumed
before.

In our analysis we have deliberately chosen the size range for small
Neptunes to be the same as that adopted by \cite{Howard:12}, to
facilitate the comparison with an interesting result they reported in
a study based on the first three quarters of \kepler\ data. They found
that small Neptunes are more common around late-type stars than
early-type stars, and that the chance, $f(T_{\rm eff})$, that a star
has a planet in the 2--4\,$R_{\earth}$ range depends roughly linearly
on its temperature. Specifically, they proposed $f(T_{\rm eff}) = f_0
+ k_T (T_{\rm eff} - 5100\,{\rm K})/1000\,{\rm K}$, valid over the
temperature range from 3600\,K to 7100\,K, with coefficients $f_0 =
0.165 \pm 0.011$ and $k_T = -0.081 \pm 0.011$.

In addition to the use in the present work of the considerably
expanded KOI list from \cite{Batalha:12} \citep[which is roughly twice
the size of the KOI list from \citealt{Borucki:11} used by][]{Howard:12},
there are a number of other significant differences between our
analysis and theirs including the fact that we account for false
positives, and we use a different model for the detection efficiency
of \kepler. It is of considerable interest, therefore, to see if their
result still holds, as it could provide important insights into the
process of planet formation and/or migration.

We first repeated our analysis as before, with no assumed dependence
of the planet frequencies on the spectral type of the host star, but
we adjusted other assumptions to match those of \cite{Howard:12}.
Instead of our modified detection model (linear ramp;
Sect.~\ref{sec:detectionmodel}), we adopted a fixed SNR threshold of
10, as they did.  Also, rather than assuming logarithmic distributions
within each of our sub-bins (which \citealt{Howard:12} also
considered), we assigned to each planet a radius equal to the value
corresponding to the center of the bin from \cite{Howard:12}, as they
did. Adopting the same linear relation $f(T_{\rm eff})$ proposed by
\cite{Howard:12}, a least-squares fit to the frequencies from our
simulations as a function of host star effective temperature yielded
the coefficients $f_0 = 0.170$ and $k_T = -0.082$, in very good
agreement with their results.

\begin{figure}
\vskip 10pt
\begin{center}
\epsscale{1.15} 
\plotone{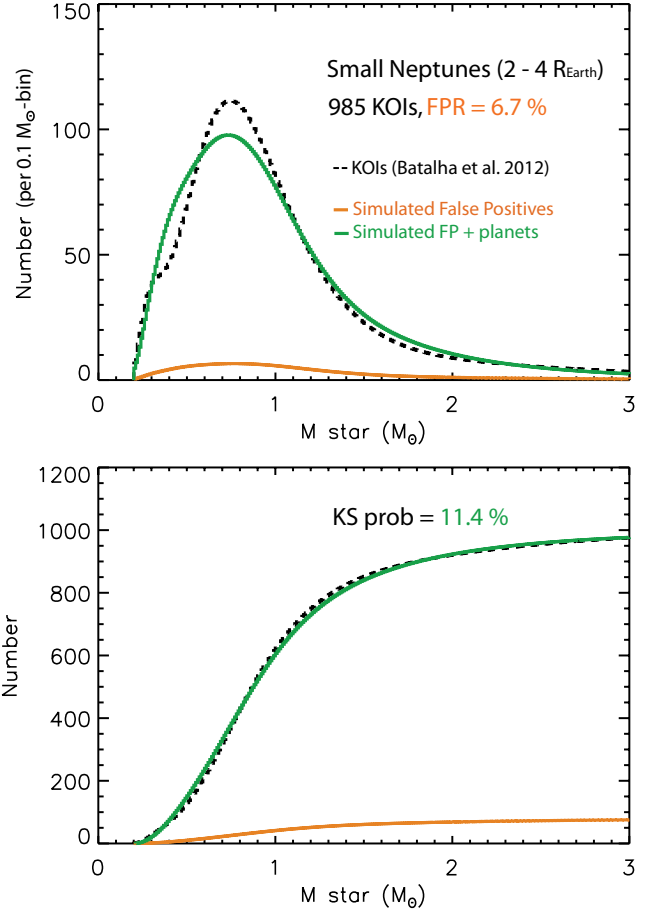}
\vskip 1pt
 \caption{Similar to Fig.~\ref{fig:mstar_6}, for the small Neptunes. A
 K-S test indicates there is no significant correlation between planet
 occurrence and stellar mass.\label{fig:mstar_2}}
\end{center}
\end{figure}

However, returning to the assumptions of our work in this paper
(linear ramp detection model, logarithmic/quadratic size distribution, and false
positive corrections), we find that the correlation between planet
occurrence and spectral type (or equivalently $T_{\rm eff}$, or mass)
for the small Neptunes all but disappears (Figure~\ref{fig:mstar_2}):
a K-S test indicates that the KOI distribution and our simulated
population of false positives and true planets (with the assumption of
no mass dependence) are not significantly different (false alarm
probability = 11.4\%). Thus we do not confirm the \cite{Howard:12}
finding.

\begin{figure}
\vskip 10pt
\begin{center}
\epsscale{1.15} 
\plotone{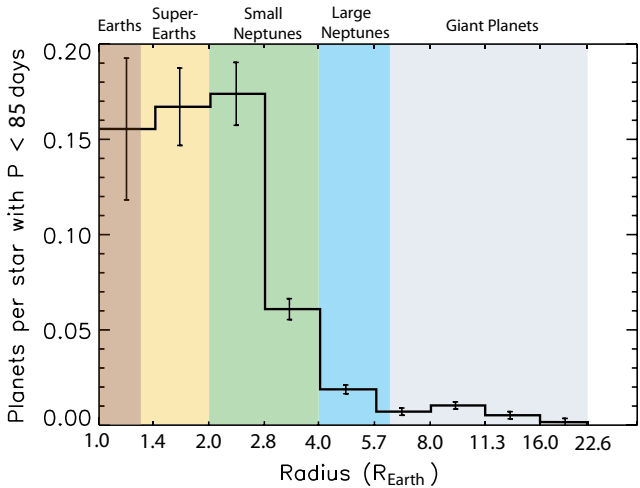}
\vskip 1pt
 \caption{Average number of planets per size bin for main sequence
 FGKM stars, determined here from the Q1--Q6 \kepler\ data and
 corrected for false positives and incompleteness.\label{fig:sizes}}
\end{center}
\end{figure}

A number of results from our simulations help to explain why we see no
dependence of the planet frequency on spectral type, whereas
\cite{Howard:12} did. One is that the median SNR for small Neptunes
(detectable or not) in our simulated sample is 12.5, a value for which
we have shown that the KOI list is likely incomplete (see
Sect.~\ref{sec:detectionmodel}). This means that a significant number
of small Neptunes in the Kepler field have not been recovered as KOIs,
especially those transiting larger stars.  Secondly, we find that the
distribution of sizes inside the small Neptune class rises sharply
towards smaller radii. This is shown in Figure~\ref{fig:sizes}. Since
these more numerous smaller planets are easier to detect around
late-type stars, this artificially boosts the occurrence of planets
around such stars.  Thirdly, the false positive rate is slightly
higher for later-type stars, again resulting in a higher planet
occurrence around those stars, if not corrected for.

\subsection{Super-Earths (1.25--2\,$R_{\earth}$)}
\label{sec:super-earths}

\begin{figure}
\vskip 10pt
\begin{center}
\epsscale{1.15} 
\plotone{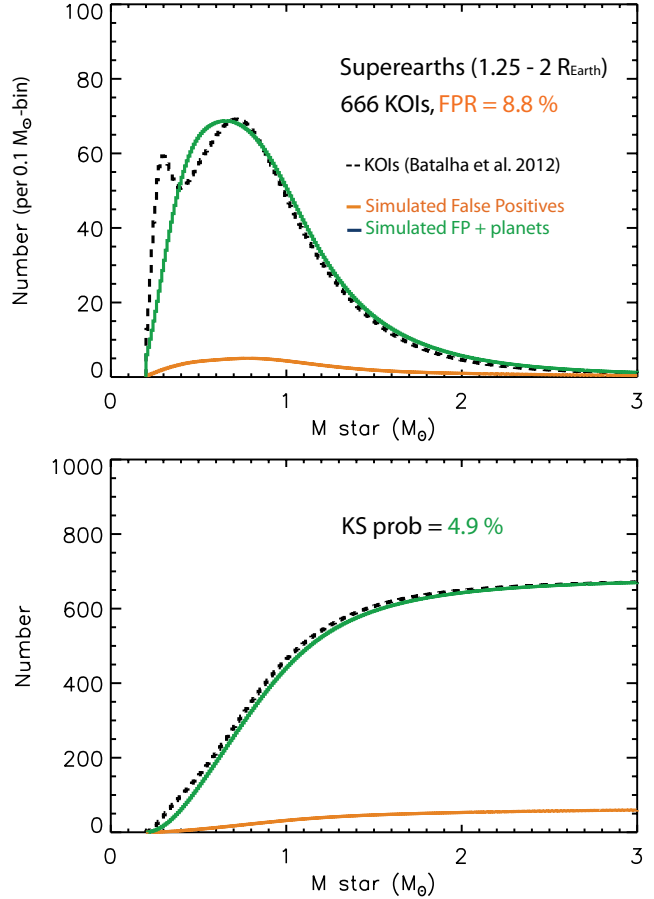}
\vskip 1pt
 \caption{Similar to Figure~\ref{fig:mstar_6}, for
 super-Earths.\label{fig:mstar_125}}
\end{center}
\end{figure}

According to our simulations the overall average number of
super-Earths per star out to periods of 145 days is close to 30\%. 
The distribution of host star masses for the super-Earths
is shown in Figure~\ref{fig:mstar_125}. While there is a hint that
planets of this size may be less common around M dwarfs than around
hotter stars, a K-S test indicates that the simulated and real
distributions are not significantly different (false alarm probability
of 4.9\%).

\subsection{Earths (0.8--1.25\,$R_{\earth}$)}
\label{sec:earths}

\begin{figure}
\vskip 10pt
\begin{center}
\epsscale{1.15}
\plotone{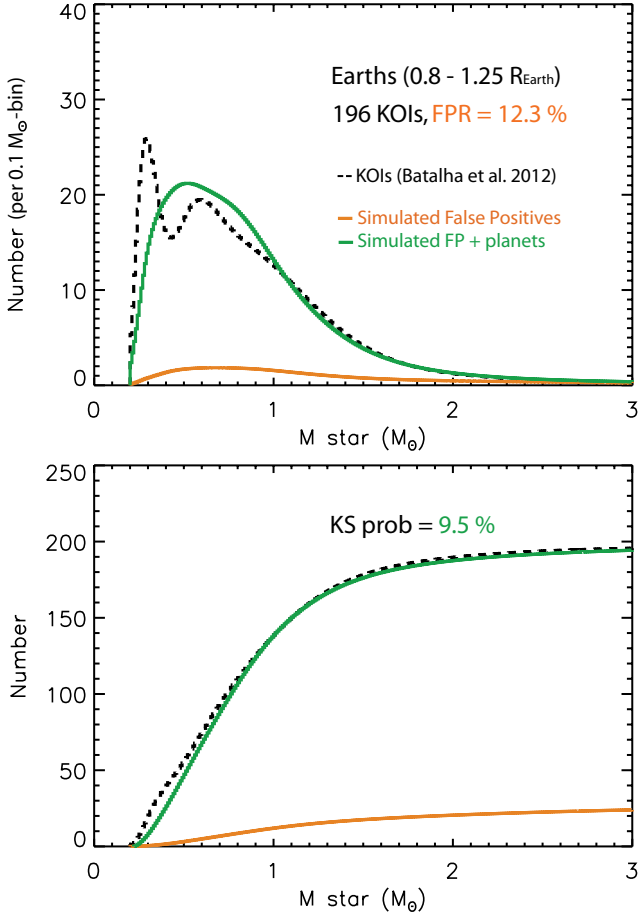}
\vskip 1pt
 \caption{Similar to Figure~\ref{fig:mstar_6}, for Earth-size
 planets.\label{fig:mstar_08}}
\end{center}
\end{figure}

As indicated in Table~\ref{tab:occurrence_period_cum}, the overall
rate of occurrence (average number of planets per star) we find for
Earth-size planets is 18.4\%, for orbital periods up to 85
days. Similarly to the case for larger planets, our simulated
population of false positives and Earth-size planets is a good match
to the KOIs in this class, without the need to invoke any dependence
on the mass of the host star (see Figure~\ref{fig:mstar_08}).

Among the Earth-size planets that we have randomly assigned to KIC
target stars in our simulations, we find that approximately 23\% have
SNRs above 7.1, but only about 10\% would be actually be detected
according to our ramp model for the \kepler\ recovery rate. These are
perhaps the most interesting objects from a scientific point of view.
Our results also indicate that 12.3\% of the Earth-size KOIs are false
positives (Table~\ref{tab:results}). This fraction is small enough to
allow statistical analyses based on the KOI sample, but is too large
to claim that any individual Earth-size KOI is a bona-fide planet
without further examination. Ruling out the possibility of a false
positive is of critical importance for the goal of confidently
detecting the first Earth-size planets in the habitable zone of their
parent star.

\begin{figure}
\begin{center}
\epsscale{1.15} 
\plotone{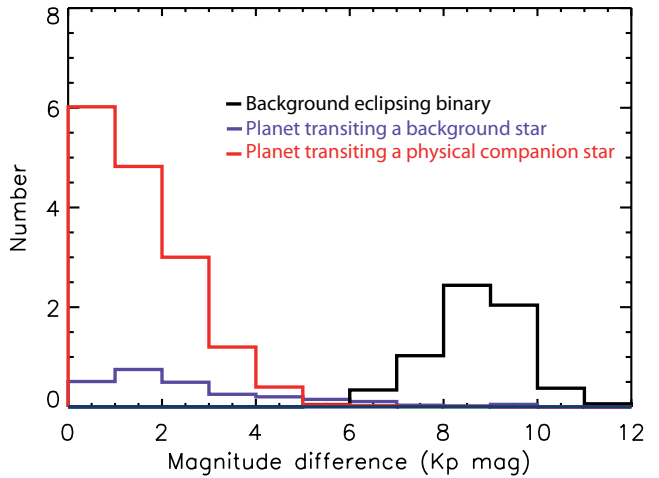}
\vskip 1pt
 \caption{Histogram of the magnitude difference (in the \kepler\ band)
relative to the target of the various kinds of false positives that
could be interpreted as a transiting Earth-size planet. Background
eclipsing binaries and larger planets transiting a physically
associated star are the main sources, with larger planets transiting
background stars being less important. Eclipsing binaries bound to the
target produce signals that are typically too deep to be confused with
an Earth.
\label{fig:dmag}}
\end{center}
\end{figure}

On the basis of our simulations we may predict the kinds of false
positives that can most easily mimic an Earth-size transit, so that
observational follow-up efforts may be better focused toward the
validation of the planetary nature of such a signal.
Figure~\ref{fig:dmag} shows a histogram of the different kinds of
false positives that result in photometric signals similar to
Earth-size transiting planets, as a function of their magnitude
difference compared to the \kepler\ target.  

There are two dominant sources of false positives for this class of
signals.  One is background eclipsing binaries, most of which are
expected to be between 8 and 10 magnitudes fainter than the \kepler\
target in the $K\!p$ passband, and some will be even fainter. The most
effective way of ruling out background eclipsing binaries is by
placing tight limits on the presence of such contaminants as a
function of angular separation from the target.  In previous planet
validations with \blender\ \citep[e.g.,][]{Fressin:11, Cochran:11,
Borucki:12, Fressin:12} the constraints from ground-based high-spatial
resolution adaptive optics imaging have played a crucial role in
excluding many background stars beyond a fraction of an arcsec from
the target.  However, these observations typically only reach
magnitude differences up to 8--9 mag \citep[e.g.,][]{Batalha:11}, and
such dim sources can only be detected at considerably larger angular
separations of several arcsec. Any closer companions of this
brightness would be missed. Since background eclipsing binaries
mimicking an Earth-size transit can be fainter still, other more
powerful space-based resources may be needed in some cases such as
choronography or imaging with HST.

Another major contributor to false positives, according to
Figure~\ref{fig:dmag}, is larger planets transiting a physically bound
companion star. In this case the angular separations from the target
are significantly smaller than for background binaries, and imaging is
of relatively little help. Nevertheless, considerable power to rule
out such blends can be gained from high-SNR spectroscopic observations
in the optical or near-infrared, which can provide useful limits on
the presence of very close companions in the form of maximum companion
brightness as a function of radial-velocity difference compared to the
target.

\section{Transit durations}
\label{sec:duration}

\begin{figure}
\begin{center}
\epsscale{1.1} 
\plotone{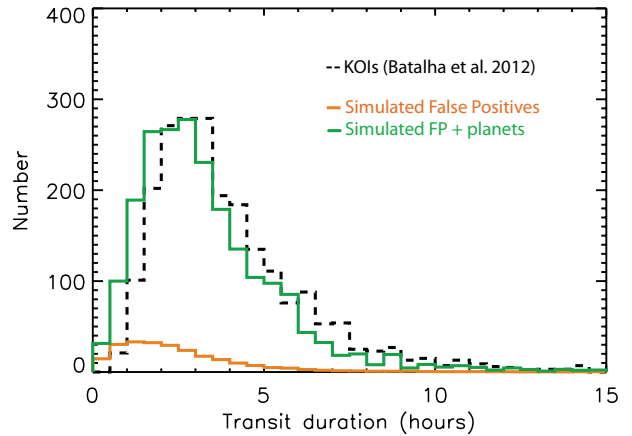}
\vskip 1pt
 \caption{Histogram of the transit durations for the total population
 of KOIs from Earth-size to giant planets (dotted black line),
 simulated false positives (orange), and the sum of the simulated
 false positives and simulated planets (green).  The excess of
 simulated false positives and planets with very short durations, in
 comparison with the KOI distribution, is mainly due to the fact that
 the \kepler\ pipeline was not designed to extract candidates with
 durations less than 1 hour.
 \label{fig:duration}}
\end{center}
\end{figure}

An additional result of interest from the present study concerns the
transit durations.  Figure~\ref{fig:duration} shows the distribution
of durations for our simulated false positives and planets (all sizes)
compared with the distribution for the KOIs from \cite{Batalha:12}.
We find an excess of short durations, some under 1 hour. This is
likely explained by the fact that the \kepler\ pipeline is only
designed, in principle, to search for transits with durations between
1 and 16 hours \citep{Jenkins:10}. These short-duration transits have
been included in our simulations because even though they were
nominally not searched for, some KOIs in the list of \cite{Batalha:12}
actually do have such short durations. More importantly, this result
suggests that there should actually be more than 100 additional
planets with such extremely short durations that may be detectable in
the light curves. Efforts to look for them may reveal an interesting
and unexplored population.

\section{Discussion}
\label{sec:discussion}

We have endeavored in this work to adopt assumptions that are as
reasonable and realistic as is practical, in order to ensure the
results are as accurate as possible. We have gone to considerable
lengths to test and adopt a sensible model for the detection
efficiency of \kepler, we have used informed estimates of quantities
such as number densities of stars, frequencies of binaries and
multiple systems, and frequencies of eclipsing binaries in the
\kepler\ field, and we have taken into account numerous other details
in our simulations that similar studies have generally not considered.
These efforts notwithstanding, unavoidable idiosyncrasies in the way
the \kepler\ photometry is handled and in the process by which the
most recent catalog of KOIs was assembled mean that it is very
difficult to avoid subtle biases when extracting information on the
false positive rate and the frequencies of planets from this somewhat
inhomogeneous data set.

One example of these difficulties is the vetting process followed by
the \kepler\ team. It is quite likely that this procedure has rejected
more false positives than we have in our simulations, especially for
the high SNR candidates, because of the application of additional
criteria based on the light curves themselves. For instance,
information on the shape of the transits, which is not explicitly used
in our work, can be extremely useful for excluding blends, as
demonstrated forcefully in a number of validation studies of \kepler\
candidates with \blender\ \citep[e.g.,][]{Cochran:11, Ballard:11,
Fressin:12b, Gautier:12, Borucki:12}. This can often reduce the false
positive frequencies by orders of magnitude. It is reasonable to
assume that interesting photometric signals have only been promoted to
KOIs by the \kepler\ team if an acceptable fit to the light curve was
possible with a transit model.  However, estimating how many signals
were rejected because of a poor fit is non-trivial, and is further
complicated when the shapes are distorted (widened) by unrecognized
transit timing variations.  In any case, this would mostly involve
cases with relatively high SNR where the shape is well defined, such
as larger planets with multiple transits. Therefore, this criterion
would generally only rule out false positives involving larger
eclipsing objects, and not blends involving other planets or very
small stars, which constitute the majority of false positive sources.

On the other hand, not all KOIs in the catalog of \cite{Batalha:12}
have been subjected to the same level of vetting. For example, the
KOIs from the earlier list by \cite{Borucki:11}, which were
incorporated into the new catalog, did not benefit from a systematic
multi-quarter centroid motion analysis as newer candidates did, and
may therefore have a somewhat higher rate of false positives.

This effect and the one described previously will tend to compensate
each other, and as a result we do not believe the false positive rates
in Table~\ref{tab:results} should be much affected, particularly since
the biases would mostly influence scenarios involving eclipsing
binaries rather than stars transited by larger planets, and the
former happen to be less numerous.

An additional source of error in our analysis comes from possible
biases in the stellar characteristics provided in the KIC, reported by
a number of authors. \cite{Pinsonneault:12} found effective
temperatures that are typically $\sim$200\,K hotter than listed in the
KIC.  \cite{Muirhead:12} reported that the masses and radii for cool
stars with $T_{\rm eff} < 4400$\,K are overestimated, a result
confirmed by \cite{Dressing:12}. The latter authors also found support
for the earlier result by \cite{Mann:12} that nearly all (93--97\%) of
the bright and cool unclassified stars in the KIC with $K\!p < 14$ and
$T_{\rm eff} < 4000$\,K are giants. The systematic errors in the KIC
stellar parameters are likely to affect the estimated planetary radii
(blurring or shifting the boundaries of the different planet classes),
and may also impact our results regarding the correlation (or lack
thereof) between the occurrence of small planets (small Neptunes,
super-Earths) and spectral type, to some extent, possibly changing the
global false positive rate and planet occurrence results.

There are also some indications of possible biases in the fitted
transit parameters reported by \cite{Batalha:12}. For example, the
median impact parameter of the entire sample is 0.706, which seems
inconsistent with an isotropic distribution of inclination angles. The
impact parameter is typically correlated with the normalized
semi-major axis $a/R_{\star}$ in the fits (related to the duration),
as well as with the transit depth ($\propto R_p^2/R_{\star}^2$).
There is the potential, therefore, for additional systematic errors in
the estimated radii for the planet candidates, and in the durations,
which may explain part of the differences noted in
Figure~\ref{fig:duration}.

Another factor that can influence the durations is the orbital
eccentricity distribution. The overall effect of increasing the
eccentricities for the simulated planets is to shift the duration
distribution slightly toward smaller values.  This can be understood
by realizing that, although transits occurring near apastron last
longer, the chance that they will happen is smaller than transits
occurring when the planet is closer to the star. This connection
between eccentricities and transit durations may in fact be exploited
to characterize the distribution of planetary eccentricities using the
durations, as done by \cite{Moorhead:11} based on an earlier release
of candidates from the first two quarters of \kepler\ observations
\citep{Borucki:11}. We have refrained from pursuing such a project
here with the updated candidate list, as we believe uncertainties in
the stellar parameters from the KIC, the transit parameters, and in
the efficiency of the \kepler\ detection pipeline for very short
durations are still too important to allow an unbiased
characterization of the eccentricity distribution.

\section{Conclusions}
\label{sec:conclusions}

The \kepler\ Mission was conceived with the objective of determining
the frequency of Earth-size and larger planets, including the ones in
or near the habitable zone of their parent stars, and determining
their properties. This eminently statistical goal requires a good
understanding of the false positive rate among planet candidates, and
of the actual detection capability of \kepler.

In this work we have developed a detailed simulation of the entire
\kepler\ transit survey based on observations from quarters 1 through
6, designed to extract information on the occurrence rates of planets
of different sizes as a function of orbital period.  In the process we
have also been able to reconstruct a model of the detection efficiency
of the global \kepler\ pipeline, and learn about the incidence of
false positives of different kinds. We have made an effort to use
assumptions in our simulations that we consider to be more realistic
than those used in earlier studies of false positives such as that of
\cite{Morton:11}.  For convenience we have classified planets into
five categories by size: giant planets (6--22\,$R_{\earth}$), large
Neptunes (4--6\,$R_{\earth}$), small Neptunes (2--4\,$R_{\earth}$),
super-Earths (1.25--2\,$R_{\earth}$), and Earths
(0.8--1.25\,$R_{\earth}$).  The main results may be summarized as
follows:
{\setlength{\leftmargini}{0pt}
\begin{enumerate}
\item We infer the rate of false positives in the \kepler\ field for
each planet class, broken down by the type of configuration of the
blend (Table~\ref{tab:results}). This includes eclipsing binaries or
other stars transited by larger planets, either of which may be in the
background or physically associated with the target. The dominant type
of false positive for all planet classes is physically associated
stars transited by a larger planet. The overall false positive rate is
17.7\% for giant planets, decreasing to a minimum of 6.7\% for small
Neptunes, and increasing again up to 12.3\% for Earth-size planets. On
average the mean false positive rate for planets of all sizes is $9.4
\pm 0.9$\%, which may be compared with the value of 4\% derived by
MJ11.  The difference is due in part to our inclusion of planets
transiting companion stars as blends, but other factors listed in
Sect.~\ref{sec:introduction} also have a significant impact.

\item We derive the occurrence rate of planets of different sizes as a
function of their orbital period. These results are presented in two
different forms: in terms of the average number of planets per star
(Table~\ref{tab:occurrence_period} for different period bins, and
cumulative rates in Table~\ref{tab:occurrence_period_cum}), and also
expressed as the percentage of stars with at least one planet
(Table~\ref{tab:stars_with_planets}). {\modif For planets larger than
2\,$R_{\earth}$ and periods up to 50 days we may compare our
occurrence rates with those of \cite{Howard:12}, which are also based
on \kepler\ candidates. Our results indicate a rate of $0.209 \pm
0.013$ planets per star, which is slightly larger than their estimate
of $0.166 \pm 0.009$. The excess is due to the previously mentioned
differences between our approaches. In particular, our prescription
for the actual detection recovery rate of \kepler\ (see
Sect.~\ref{sec:detectionmodel}) leads to a larger occurrence of small
Neptunes.}  For Earth-size planets we find that about 16.5\% of stars
have at least one planet in this category with orbital periods up to
85 days, beyond which the statistics are still too poor to provide
results. Rates for other planet sizes are given in
Table~\ref{tab:stars_with_planets}. The percentage of stars that have
at least one planet of any size out to 85 days is approximately
52\%. This high percentage is broadly in agreement with results from
the HARPS radial velocity survey of \cite{Mayor:11}.  Those authors
reported that the rate of low-mass planets (having masses {\modif
between 1 and 30\,$M_{\earth}$}) with periods shorter than 100 days is
larger than 50\%. While the figures are similar, we must keep in mind
that they refer to two different planet properties (radii and masses).
A relevant improvement of our procedures we plan for the future is to
incorporate a model of the global mass-radius distribution of close-in
planets that would simultaneously enable a good match to the mass
distribution from Doppler surveys and the radius distribution from
\kepler.

\item We find that the effective detection efficiency of
\kepler\ differs from that expected from the nominal signal-to-noise
criterion applied during the Mission, namely, that 50\% of signals with
SNR greater than 7.1 are detectable. Instead, we find that the actual
distribution of SNRs for the KOIs released by \cite{Batalha:12} is
better represented with a detection efficiency that increases linearly
from 0\% for a SNR of 6 to 100\% for a SNR of 16.

\item After accounting for false positives and the effective
detection efficiency of \kepler\ as described above, we find no
significant dependence of the rates of occurrence as a function of the
spectral type (or mass, or temperature) of the host star. This
contrasts with the findings by \cite{Howard:12}, who found that for
the small Neptunes (2--4\,$R_{\earth}$) M stars have higher planet
frequencies than F stars.

\item We find an apparent excess of transits of very short
duration (less than one hour). Such transits have not explicitly been
looked for in the \kepler\ pipeline.

\end{enumerate}

The planet occurrence rates provided in
Table~\ref{tab:occurrence_period} should be useful in future planet
validation studies (e.g., using the \blender\ procedure) to estimate
the ``planet prior'' (\emph{a priori} chance of a planet) for a
candidate of a given size and period. A comparison of this prior with
the \emph{a priori} chance that the candidate is a false positive,
incorporating constraints from any available follow-up observations,
could then be used to establish the confidence level of the
validation.

Our technique provides an estimate of the occurrence of planets
orbiting dwarf stars in the solar neighborhood that is based almost
entirely on \kepler\ observations and modeling. Improvements in the
\kepler\ pipeline, in the understanding of the detection efficiency
and with the addition of more quarters of data, will allow to improve
the planet occurrence estimates, extend them to longer periods, and
study their relations with their host stars characteristics.

\acknowledgements

We thank the anonymous referee for helpful suggestions on the original
manuscript. GT acknowledges partial support for this work from NASA
grant NNX12AC75G (\kepler\ Participating Scientist Program).

{\it Facilities:} \facility{\kepler\ Mission}.


\clearpage

\section*{Appendix: Modeling of the transit SNR and of the \Kepler\
photo-centroid shift.}

\subsection{Signal-to-noise calculation}
\label{sec:snr}

In this work we have adopted the use of the Combined Differential
Photometric Precision (CDPP), which is an empirical measure of the
effective noise seen by transits as a function of their duration
\citep{Jenkins:10b,Christiansen:12}. The CDPP is obtained as a time series
for each star for each of 14 trial transit durations ranging from 1.5
hours to 15 hours as a by-product of the search for transiting planets
by the Kepler Transiting Planet Search (TPS) pipeline. TPS
characterizes the Power Spectral Density (PSD) of each observed flux
time series and calculates the expected SNR of the reference transit
pulse at each time step. The CDPP time series are obtained by dividing
the reference transit depth by the SNR time series, thereby allowing
the SNR to be calculated easily for any depth transit of the given
duration. We use the \emph{rms} CDPP calculated across each quarter of
Kepler observations, and take the median value for each star across Q1
through Q6, interpolating across the 14 CDPP transit durations to
estimate the CDPP for each simulated transit or false positive eclipse
duration.

The measured CDPP is empirical and accounts
for the three known sources of noise: Poisson errors from the number
of photons received, which depends on star brightness, the stellar
variability noise due to stellar surface physics including spots,
turbulence (e.g., granulation), acoustic $p$-modes, and magnetic
effects, and the residual instrumental effects.

The signal-to-noise ratio of a transit is defined as
\begin{equation}\label{eq:snr}
{\rm SNR} = \frac{\delta}{{\rm CDPP}_{\rm eff}} \sqrt{\frac{t_{\rm
obs} f}{P}}~,
\end{equation}
where $\delta$ is the photometric depth of the signal and is computed
as $\delta=R_p^2/R_{\star}^2$ for a transiting planet of radius $R_p$
transiting a star of radius $R_{\star}$, or as
\begin{equation}
\delta = \frac{R_{\rm ecl}^2}{R_{\rm blend}^2} \frac{F_{\rm blend}}{F_{\rm blend}
+F_{\rm KIC}}
\end{equation}
for a blend involving an object eclipsing a blending star in the
photometric aperture of the KIC target. In the above expression
$F_{\rm blend}/(F_{\rm blend}+F_{\rm KIC})$ is the contribution in the
\kepler\ bandpass of the flux of the blending star in the photometric
aperture normalized to the sum of the blending star and KIC target
fluxes. The symbol $t_{\rm obs}$ is the duration of the \kepler\
observations from Q1 to Q6, $f$ is target-specific fraction of the
total time the target was observed, and $P$ is the orbital period of
the transiting object. The transit duration depends on the mass, size,
and period of the eclipsing object, all of which are known from our
simulations. Assumptions on the eccentricity have some impact on the
SNR through the duration. However, the duration enters only as the
square root, so any errors are reduced by a factor of two.

\subsection{Centroid shift constraint}

The most useful observational constraint available to rule out false
positives that does not require additional observations is obtained by
measuring the photocenter displacement during the transit. If the
transit signal is due to a diluted eclipse of another star in the same
photometric aperture as the target, there will generally be a shift in
the position of the photocenter that occurs during the transit, as the
neighboring star contributes less of the total flux at those times.

Of the $\sim$2300 KOIs in the cumulative catalog by \cite{Batalha:12},
the 1023 new ones that were added to the prior list from
\cite{Borucki:11} were subjected to a multi-quarter centroid motion
analysis by the \kepler\ team. This analysis provides a maximum
angular separation that corresponds to a 3-$\sigma$ limit beyond which
a false positive would have been identified.

In their study \cite{Morton:11} assumed that this exclusion radius
scales linearly with the flux from the star and the transit depth,
with a lower limit they set at 2\arcsec\ and an upper limit at
6\farcs4. The sample of 1023 new KOIs with multi-quarter centroid
analysis enables us to re-examine the model of MJ11, and test the
strength of the proposed correlations between the centroid exclusion
radius and those two parameters (stellar magnitude and transit
depth). We also studied the dependencies of the centroid exclusion
radius with several other characteristics related to the star and the
transit detection:
\begin{itemize}
\item {\bf Spectral type.} By virtue of their different age, levels of
activity, and rotation periods, stars of different spectral types may
show different noise patterns that can impact the centroid exclusion
radius;

\item {\bf Galactic latitude.} \kepler\ targets located closer to the
Galactic plane are likely to have more contamination from background
stars in their aperture, which could directly impact the centroid
exclusion radius;

\item {\bf Noise level.} We investigated whether the centroid
exclusion radius is correlated with the CDPP of each KOI;

\item {\bf Transit signal-to-noise ratio.} As described in the first
part of this Appendix, this parameter is correlated with both the CDPP
and the transit depth, along with the number of transits and the
transit duration.
\end{itemize}

For this paper we require a prescription for predicting the
approximate multi-quarter centroid exclusion radius for each \kepler\
target to be used in our simulations.  It is sufficient for our
purposes to be able to predict a reasonable \emph{range} for this
quantity.  Figure~\ref{fig:centroid} shows that the centroid exclusion
radius has a large scatter, regardless of which parameter we display
it against.  Median values in appropriate bins do not appear to show
any correlation with the spectral type (or equivalently stellar mass
$M_{\star}$), Galactic latitude, or the CDPP. The centroid exclusion
radius shows only a weak correlation with the \kepler\ magnitude, but
different from the linear correlation with the flux proposed by MJ11:
there is little variation except for the faintest bin ($K\!p > 15.8$
mag), which is likely due to the higher background noise level, and
for the brightest stars ($K\!p < 12$ mag), due to the fact that
\kepler\ stars saturate below a magnitude of about 11.5
\citep{Batalha:12}.  The clearest correlation in the centroid
exclusion radius is with the transit depth $(R_p/R_{\star})^2$ (with a
Pearson correlation coefficient of $-0.1$), and with the transit SNR
(with a Pearson correlation coefficient of $-0.08$), which is of
course highly correlated with the transit depth (Eq.\,[\ref{eq:snr}]).

In order to make use of this correlation with the transit SNR for our
simulations, and at the same time to account for the large scatter
present in that correlation, we proceeded as follows.  We first
computed the SNR of each false positive scenario we simulated using
Eq.\,[\ref{eq:snr}], and we then selected a random exclusion radius
from the sub-sample of the 1023 KOIs having a SNR within 10\% of the
one of this false positive. This is the exclusion radius we used for
our emulation of the \kepler\ vetting procedure.

\begin{figure}[b!]
\vskip 10pt
\begin{center}
\epsscale{1.15} 
\plotone{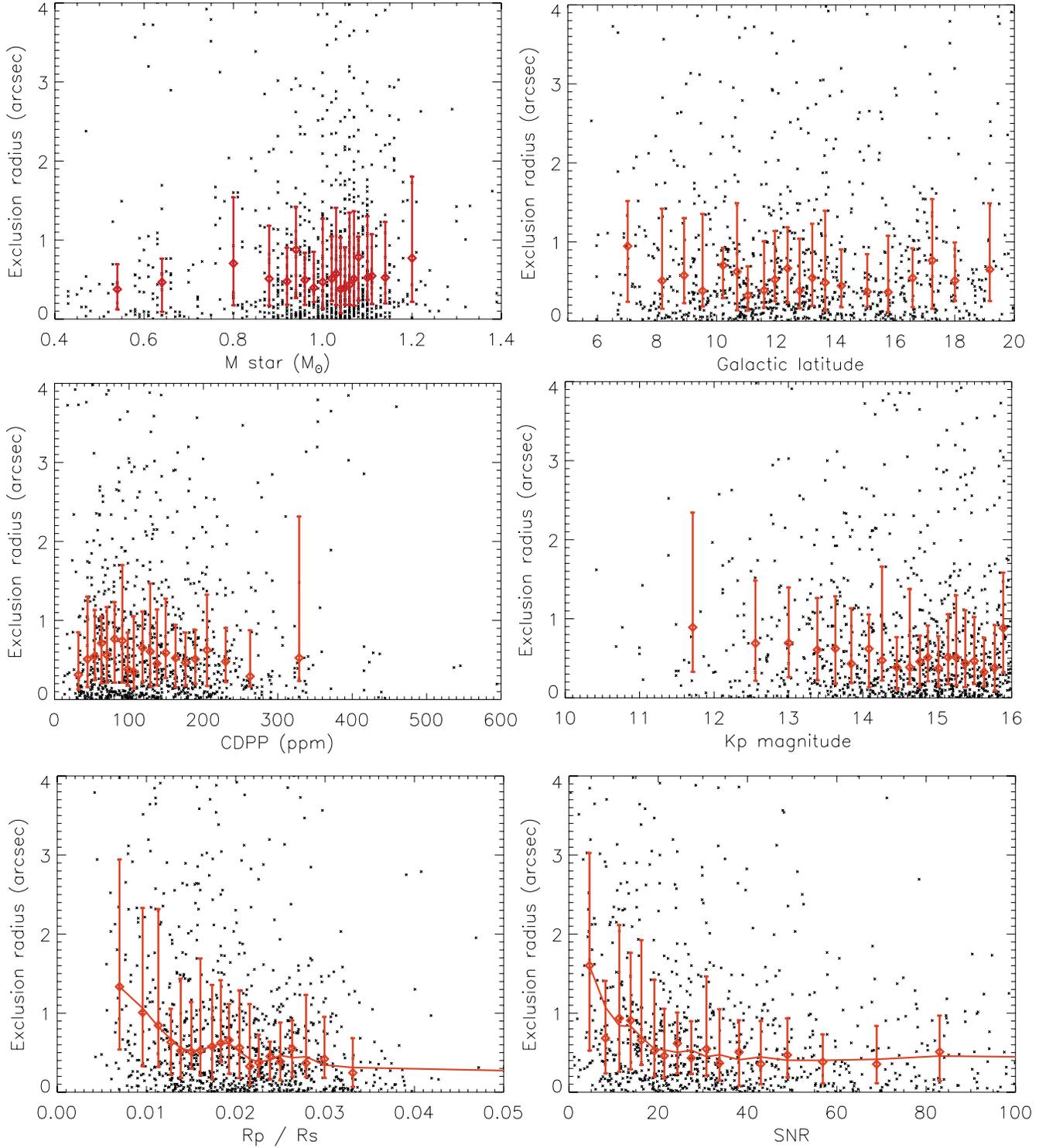}
%

\caption{Distribution of the centroid exclusion radius for the 1023
KOIs for which a multi-quarter centroid analysis is available. Black
dots are the individual values for the 1023 KOIs. The red diamonds are
the median values for bins of 50 KOIs ranked as a function of
parameter represented on the horizontal axis. The error bars
correspond to the 25 and 75 percentiles. The centroid exclusion radius
shows no significant correlation with the spectral type (stellar
mass), Galactic latitude, noise level (CDPP), or $K\!p$ magnitude of
the KOIs. The only significant correlation is with the radius ratio
($R_p/R_{\star}$) and the transit SNR (which is correlated with
$R_p/R_{\star}$). The red line shows the sliding median of the
centroid exclusion radius as a function of these two
parameters.\label{fig:centroid}}

\end{center}
\end{figure}



\begin{thebibliography}

\bibitem[Ballard et al.(2011)]{Ballard:11} Ballard, S., Fabrycky, 
D., Fressin, F., et al.\ 2011, \apj, 743, 200 

\bibitem[Batalha et al.(2011)]{Batalha:11} 
 Batalha, N.\ M.\ et al.\ 2011, \apj, 729, 27

\bibitem[Batalha et al.(2012)]{Batalha:12} Batalha, N.~M., Rowe, 
J.~F., Bryson, S.~T., et al.\ 2012, arXiv:1202.5852 

\bibitem[Borucki et al.(2011)]{Borucki:11} Borucki, W.~J., Koch, 
D.~G., Basri, G., et al.\ 2011, \apj, 736, 19  

\bibitem[Borucki et al.(2012)]{Borucki:12} Borucki, W.~J., Koch, 
D.~G., Batalha, N., et al.\ 2012, \apj, 745, 120 

\bibitem[Bowler et al.(2010)]{Bowler:10} Bowler, B.~P., Johnson, 
J.~A., Marcy, G.~W., et al.\ 2010, \apj, 709, 396 

\bibitem[Brown(2003)]{Brown:03} Brown, T.~M.\ 2003, \apjl, 593, 
L125 

\bibitem[Brown et al.(2011)]{Brown:11}
 Brown, T.\ M., Latham, D.\ W., Everett, M.\ E., \& Esquerdo, G.\ A.,
 \aj, 142, 112, 2011

\bibitem[Burke et al.(2012)]{Burke:12} Burke, C.~J., 
Christiansen, J.~L., Jenkins, J.~M., et al.\ 2012, American Astronomical 
Society Meeting Abstracts, 219, \#245.07 

\bibitem[Catanzarite \& Shao(2011)]{Catanzarite:11} Catanzarite, J., \& Shao, M.\ 2011, \apj, 738, 151 

\bibitem[Charbonneau et al.(2009)]{Charbonneau:09} Charbonneau, D.\ et al.\ 2009, \nat, 462, 891

\bibitem[Christiansen et al.(2012)]{Christiansen:12} Christiansen, 
J.~L., Jenkins, J.~M., Barclay, T.~S., et al.\ 2012, arXiv:1208.0595

\bibitem[Cochran et al.(2011)]{Cochran:11} Cochran, W.~D., 
Fabrycky, D.~C., Torres, G., et al.\ 2011, \apjs, 197, 7 
 
\bibitem[Demory \& Seager(2011)]{Demory:11} Demory, B.-O., \& Seager, S.\ 2011, \apjs, 197, 12 


\bibitem[Dotter et al.(2008)]{Dotter:08} Dotter, A., Chaboyer, 
B., Jevremovi{\'c}, D., et al.\ 2008, \apjs, 178, 89

\bibitem[Dressing et al.(2012)]{Dressing:12} Dressing, C., et al. submitted to \apj 

\bibitem[Farmer \& Agol(2003)]{Farmer:03} Farmer, A.~J., \& Agol, E.\ 2003, \apj, 592, 1151 

\bibitem[Fressin et al.(2011)]{Fressin:11} Fressin, F., Torres, 
G., D{\'e}sert, J.-M., et al.\ 2011, \apjs, 197, 5

\bibitem[Fressin et al.(2012)]{Fressin:12} Fressin, F., Torres, 
G., Rowe, J.~F., et al.\ 2012, \nat, 482, 195 

\bibitem[Fressin et al.(2012)]{Fressin:12b} Fressin, F., Torres, 
G., Pont, F., et al.\ 2012, \apj, 745, 81 

\bibitem[Gautier et al.(2012)]{Gautier:12} Gautier, T.~N., III, 
Charbonneau, D., Rowe, J.~F., et al.\ 2012, \apj, 749, 15 

\bibitem[Girardi et al.(2000)]{Girardi:00}
 Girardi, L., Bressan, A., Bertelli, G., \& Chiosi, C. 2000, \aaps,
141, 371

\bibitem[Girardi et al.(2005)]{Girardi:05} Girardi, L., Groenewegen, M.~A.~T., Hatziminaoglou, E., \& da Costa, L.\ 2005, \aap, 436, 895 

\bibitem[Holman 
\& Wiegert(1999)]{Holman:99} Holman, M.~J., \& Wiegert, P.~A.\ 1999, \aj, 117, 621

\bibitem[Howard et al.(2010)]{Howard:10} Howard, A.~W., Marcy, 
G.~W., Johnson, J.~A., et al.\ 2010, Science, 330, 653

\bibitem[Howard et al.(2012)]{Howard:12} Howard, A.~W., Marcy, 
G.~W., Bryson, S.~T., et al.\ 2012, \apjs, 201, 15 

\bibitem[Jenkins et al.(1996)]{Jenkins:96} Jenkins, J.~M., Doyle, 
L.~R., \& Cullers, D.~K.\ 1996, Icarus, 119, 244

\bibitem[Jenkins et al.(2010)]{Jenkins:10} Jenkins, J.~M., 
Caldwell, D.~A., Chandrasekaran, H., et al.\ 2010, \apjl, 713, L87

\bibitem[Jenkins et al.(2010)]{Jenkins:10b} Jenkins, J.~M., 
Chandrasekaran, H., McCauliff, S.~D., et al.\ 2010, \procspie, 7740,

\bibitem[Johnson et al.(2010)]{Johnson:10} Johnson, J.~A., Aller, 
K.~M., Howard, A.~W., \& Crepp, J.~R.\ 2010, \pasp, 122, 905 

\bibitem[Latham et al.(2009)]{Latham:09} Latham, D.~W., Bakos, 
G.~{\'A}., Torres, G., et al.\ 2009, \apj, 704, 1107 

\bibitem[Lissauer et al.(2011)]{Lissauer:11}
 Lissauer, J.\ J.\ et al.\ 2011, \nat, 470, 53

\bibitem[Lissauer et al.(2012)]{Lissauer:12} Lissauer, J.~J., 
Marcy, G.~W., Rowe, J.~F., et al.\ 2012, \apj, 750, 112

\bibitem[Mann et al.(2012)]{Mann:12} Mann, A.\ W., Gaidos, E.,
L\'epine, S., \& Hilton, E.\ J. 2012, \apj, 753, 90

\bibitem[Mayor et al.(2011)]{Mayor:11} Mayor, M., Marmier, M., 
Lovis, C., et al.\ 2011, arXiv:1109.2497

\bibitem[Moorhead et al.(2011)]{Moorhead:11} Moorhead, A.~V., Ford, 
E.~B., Morehead, R.~C., et al.\ 2011, \apjs, 197, 1 

\bibitem[Morton \& Johnson(2011)]{Morton:11} Morton, T.~D., \& Johnson, J.~A.\ 2011, \apj, 738, 170 

\bibitem[Morton(2012)]{Morton:12} Morton, T.~D.\ 2012, 
arXiv:1206.1568 

\bibitem[Muirhead et al.(2012)]{Muirhead:12} Muirhead, P.~S., 
Hamren, K., Schlawin, E., et al.\ 2012, \apjl, 750, L37 

\bibitem[Pinsonneault et al.(2012)]{Pinsonneault:12} Pinsonneault, 
M.~H., An, D., Molenda-{\.Z}akowicz, J., et al.\ 2012, \apjs, 199, 30 

\bibitem[Pont et al.(2006)]{Pont:06} Pont, F., Zucker, S., 
\& Queloz, D.\ 2006, \mnras, 373, 231 

\bibitem[Pr{\v s}a et al.(2011)]{Prsa:11} Pr{\v s}a, A., 
Batalha, N., Slawson, R.~W., et al.\ 2011, \aj, 141, 83 

\bibitem[Raghavan et al.(2010)]{Raghavan:10} Raghavan, D., 
McAlister, H.~A., Henry, T.~J., et al.\ 2010, \apjs, 190, 1 

\bibitem[Robin et al.(2003)]{Robin:03}
 Robin, A.\ C., Reyl\'e, C., Derri\'ere, S., \& Picaud, S. 2003, \aap,
 409, 523

\bibitem[Rowe et al.(2010)]{Rowe:10} Rowe, J.~F., Borucki, 
W.~J., Koch, D., et al.\ 2010, \apjl, 713, L150 

\bibitem[Santerne et al.(2012)]{Santerne:12} Santerne, A., 
Díaz, R.~F., Moutou, C., et al.\ 2012, Submitted to \aap

\bibitem[Slawson et al.(2011)]{Slawson:11} Slawson, R.~W., Pr{\v 
s}a, A., Welsh, W.~F., et al.\ 2011, \aj, 142, 160

\bibitem[Torres et al.(2004)]{Torres:04}
 Torres, G., Konacki, M., Sasselov, D.\ D., \& Jha, S. 2004, \apj,
614, 979

\bibitem[Torres et al.(2011)]{Torres:11} 
 Torres, G.\ et al.\ 2011, \apj, 727, 24 

\bibitem[Traub(2012)]{Traub:12} Traub, W.~A.\ 2012, \apj, 745, 
20 

\bibitem[Youdin(2011)]{Youdin:11} Youdin, A.~N.\ 2011, \apj, 742, 
38 















































\clearpage


\begin{deluxetable}{lcccccccc}
\tabletypesize{\scriptsize}
\tablewidth{0pc}
\tablecaption{False positives for the different planet size classes.\label{tab:results}}
\tablehead{
& & \multicolumn{2}{c}{Involving Eclipsing Binaries}
& \multicolumn{2}{c}{Involving Transiting Planets}&&& \\ [+1ex]
\cline{3-4} \cline{5-6} \\ [-1.5ex]
\colhead{Class} &
\colhead{Size range} &
\colhead{Background} &
\colhead{Bound} &
\colhead{Background} &
\colhead{Bound} &
\colhead{Total} &
\colhead{KOIs} &
\colhead{FP rate}
}
\startdata
Giants & 6--22\,$R_{\earth}$ & $8.0 \pm 3.1$ & $4.7 \pm 2.2$ & $2.3 \pm 1.7$ & $24.5 \pm 5.0$ & $39.5 \pm 6.4$ & $223$ & $17.7 \pm 2.9\,\%$ \\
Large Neptunes & 4--6\,$R_{\earth}$        & $5.5 \pm 2.6$ & $0.4 \pm 0.4$ & $1.4 \pm 1.3$ & $14.4 \pm 3.8$ & $21.7 \pm 4.8$ & $137$ & $15.9 \pm 3.5\,\%$ \\
Small Neptunes & 2--4\,$R_{\earth}$   & $15.1 \pm 4.3$  & $0.2 \pm 0.2$ & $4.2 \pm 2.2$ & $46.6 \pm 9.6$ & $66.1 \pm 10.7$ & $985$ & $6.7 \pm 1.1\,\%$ \\
Super-Earths & 1.25--2\,$R_{\earth}$ & $10.5 \pm 3.6$  & $0.05 \pm 0.05$ & $3.6 \pm 2.1$ & $44.1 \pm 12.3$ & $58.3 \pm 13.0$ & $666$ & $8.8 \pm 1.9\,\%$ \\
Earths & 0.8--1.25\,$R_{\earth}$     & $5.5 \pm 2.6$ & $0.002 \pm 0.002$ & $2.2 \pm 1.6$ & $16.3 \pm 5.0$ & $24.1 \pm 5.9$ & $196$ & $12.3 \pm 3.0\,\%$ \\ [+1.5ex]
Total & 													  & $44.7 \pm 7.3$ & $5.3 \pm 2.2$ & $13.8 \pm 4.1$ & $146.0 \pm 17.5$ & $209.8 \pm 19.6$ & $2222$ & $9.4 \pm 0.9\,\%$ \\ [-2.5ex]
\enddata
%
\tablecomments{The number of KOIs in the penultimate column
corresponds to stars not considered here not to be giants \citep[$\log
g > 3.6$;][]{Brown:11}. Additionally, we have excluded the few KOIs
outside of the radius ranges of interest here (0.8--22\,$R_{\earth}$).}
\end{deluxetable}



%
%
%
%


\begin{deluxetable}{cccccccccccc}
\tabletypesize{\scriptsize} 
\tablewidth{0pc}
\tablecaption{Average number of planets per star per period bin (in percent).}
\tablehead{
\colhead{Class} &
\multicolumn{10}{c}{Period range (days)} \\ [+1.5ex]
 & 0.8-- & 2.0-- & 3.4-- & 5.9-- & 10-- & 17-- & 29-- & 50-- & 85-- & 145-- & 245--  \\ 
 & 2.0 & 3.4 & 5.9 & 10 & 17 & 29 & 50 & 85 & 145 & 245 & 418\tablenotemark{*}  
}
\startdata
Giants & 0.015  & 0.067 & 0.17  & 0.18  & 0.27  & 0.23  & 0.35  & 0.71  & 1.25  & 0.94  & 1.05 \\
& $\pm$ 0.007  & $\pm$ 0.018   & $\pm$ 0.03   & $\pm$ 0.04   & $\pm$ 0.06   & $\pm$ 0.06   & $\pm$ 0.10    & $\pm$
0.17    & $\pm$ 0.29    & $\pm$ 0.28    & $\pm$ 0.30 \\
Large Neptunes & 0.004  & 0.006 & 0.11  & 0.091 & 0.29  & 0.32  & 0.49  & 0.66  & 0.43  & 0.53  & 0.24 \\
& $\pm$ 0.003  & $\pm$ 0.006  & $\pm$ 0.03   & $\pm$ 0.030   & $\pm$ 0.07 & $\pm$ 0.08     & $\pm$ 0.12    & $\pm$
0.16    & $\pm$ 0.17    & $\pm$ 0.21    & $\pm$ 0.15 \\
Small Neptunes & 0.035  & 0.18  & 0.73  & 1.93  & 3.67  & 5.29  & 6.45  & 5.25  & 4.31  & 3.09  & \nodata \\
& $\pm$ 0.011   & $\pm$ 0.03   & $\pm$ 0.09   & $\pm$ 0.19    & $\pm$ 0.39    & $\pm$ 0.64    & $\pm$ 1.01    & $\pm$
1.05    & $\pm$ 1.03    & $\pm$ 0.90    & \\
Superearths & 0.17      & 0.74  & 1.49  & 2.90  & 4.30  & 4.49  & 5.29  & 3.66  & 6.54  & \nodata & \nodata \\
& $\pm$ 0.03   & $\pm$ 0.13    & $\pm$ 0.23    & $\pm$ 0.56    & $\pm$ 0.73    & $\pm$ 1.00    & $\pm$ 1.48    & $\pm$
1.21    & $\pm$ 2.20    & \nodata & \nodata \\
Earths & 0.18   & 0.61  & 1.72  & 2.70  & 2.70  & 2.93  & 4.08  & 3.46  & \nodata & \nodata      & \nodata     \\
& $\pm$ 0.04   & $\pm$ 0.15    & $\pm$ 0.43    & $\pm$ 0.60    & $\pm$ 0.83    & $\pm$ 1.05    & $\pm$ 1.88    & $\pm$
2.81    & & & \\ [+1.5ex]
Total  & 0.41   & 1.60  & 4.22  & 7.79  & 11.2  & 13.3  & 16.7  & 13.7  & \nodata & \nodata      & \nodata     \\
       & $\pm$ 0.05    & $\pm$ 0.20  & $\pm$ 0.50  & $\pm$ 0.85  & $\pm$ 1.2  & $\pm$ 1.6  & $\pm$ 2.6  & $\pm$
3.2  & & & \\  [-2.5ex]
\enddata 
%
\tablecomments{Planet occurrence per period bin for each class of
planet defined in Sect.~\ref{sec:classes}. The first line in each
group represents planet occurrences (in percent), and the second line
gives their error bars. The increase in the uncertainties relative to
the occurrence rates towards longer periods is due to the drop in the
geometric transit probability and detection rate. Empty fields for the
smallest planets occur where the \kepler\ results based on 6 quarters
of data are insufficient to provide an
estimate.\label{tab:occurrence_period}}
%
\tablenotetext{*}{For planets with long periods we have assumed that
two transits are sufficient for a detection. We have also corrected
the planet occurrences for periods longer than half the total duration
of the Q1--Q6 survey (670 days), to account for the fact that a
fraction of these long period planets would have shown a single
transit in the Q1--Q6 survey, depending on the transit date.}
%
\end{deluxetable}


%
%
%



\begin{deluxetable}{cccccccccccc}
\tabletypesize{\scriptsize} 
\tablewidth{0pc}
\tablecaption{Average number of planets per star for different period ranges (in percent)}
\tablehead{
\colhead{Class} &
\multicolumn{10}{c}{Period range (days)} \\ [+1.5ex]
 & 0.8-- & 0.8-- & 0.8-- & 0.8-- & 0.8-- & 0.8-- & 0.8-- & 0.8-- & 0.8-- & 0.8-- & 0.8-- \\
 & 2.0 & 3.4 & 5.9 & 10 & 17 & 29 & 50 & 85 & 145 & 245 & 418 
}
\startdata
Giants & 0.015  & 0.082 & 0.25  & 0.43  & 0.70  & 0.93  & 1.29  & 2.00  & 3.24  & 4.19  & 5.24 \\
& $\pm$ 0.007  & $\pm$ 0.019   & $\pm$ 0.04   & $\pm$ 0.05   & $\pm$ 0.08   & $\pm$ 0.10    & $\pm$ 0.14    & $\pm$
0.22    & $\pm$ 0.37    & $\pm$ 0.46    & $\pm$ 0.55 \\
Large Neptunes & 0.004  & 0.010 & 0.12  & 0.21  & 0.50  & 0.82  & 1.31  & 1.97  & 2.41  & 2.94  & 3.18 \\
& $\pm$ 0.003  & $\pm$ 0.007  & $\pm$ 0.03   & $\pm$ 0.04   & $\pm$ 0.08 & $\pm$ 0.11      & $\pm$ 0.17    & $\pm$
0.23    & $\pm$ 0.29    & $\pm$ 0.36    & $\pm$ 0.39 \\
Small Neptunes & 0.035  & 0.22  & 0.95  & 2.88  & 6.55  & 11.8  & 18.3  & 23.5  & 27.8  & 30.9  & \nodata \\
& $\pm$ 0.011   & $\pm$ 0.03   & $\pm$ 0.10  & $\pm$ 0.21    & $\pm$ 0.44    & $\pm$ 0.8    & $\pm$ 1.3    & $\pm$
1.6    & $\pm$ 1.9    & $\pm$ 2.1    & \nodata \\
Superearths & 0.17      & 0.91  & 2.40  & 5.30  & 9.60  & 14.1  & 19.4  & 23.0  & 29.6  & \nodata & \nodata \\
& $\pm$ 0.03   & $\pm$ 0.13    & $\pm$ 0.27    & $\pm$ 0.62    & $\pm$ 0.96    & $\pm$ 1.4    & $\pm$ 2.0    & $\pm$
2.4    & $\pm$ 3.2    & \nodata & \nodata \\
Earths & 0.18   & 0.79  & 2.51  & 5.21  & 7.91  & 10.8  & 14.9  & 18.4  & \nodata & \nodata & \nodata \\
& $\pm$ 0.04   & $\pm$ 0.16    & $\pm$ 0.46    & $\pm$ 0.76    & $\pm$ 1.13    & $\pm$ 1.5    & $\pm$ 2.4    & $\pm$
3.7    & \nodata & \nodata & \nodata \\ [+1.5ex]
Total & 0.41  & 2.0     & 6.2   & 14.0  & 25.3  & 38.5  & 55.2  & 68.9  & \nodata & \nodata & \nodata \\
& $\pm$ 0.05   & $\pm$ 0.2    & $\pm$ 0.5    & $\pm$ 1.0    & $\pm$ 1.6    & $\pm$ 2.2    & $\pm$ 3.4    & $\pm$
4.7    & \nodata & \nodata & \nodata \\ [-2.5ex]
\enddata 
%
\tablecomments{Cumulative planet occurrence rates in each period bin
for the planet classes defined in Sect.~\ref{sec:blends}. The top line
for each group is the cumulative occurrence rate (in percent), and the
bottom line corresponds to the
uncertainty. \label{tab:occurrence_period_cum}}
%
\end{deluxetable}


\clearpage
%

\clearpage
\begin{deluxetable}{cccccccccccc}
\tabletypesize{\scriptsize} 
\tablewidth{0pc}
\tablecaption{Percentage of stars with at least one planet for different period ranges}
\tablehead{
\colhead{Class} &
\multicolumn{10}{c}{Period range (days)} \\ [+1.5ex]
 & 0.8-- & 0.8-- & 0.8-- & 0.8-- & 0.8-- & 0.8-- & 0.8-- & 0.8-- & 0.8-- & 0.8-- & 0.8-- \\
 & 2.0 & 3.4 & 5.9 & 10 & 17 & 29 & 50 & 85 & 145 & 245 & 418 
}
\startdata
Giants & 0.015	& 0.082	& 0.25	& 0.43	& 0.70	& 0.93	& 1.29	& 1.97	& 3.19	& 4.09	& 5.12 \\
& $\pm$ 0.007	& $\pm$ 0.019	& $\pm$ 0.04	& $\pm$ 0.05	& $\pm$ 0.08	& $\pm$ 0.10	& $\pm$ 0.14	& $\pm$ 0.22	& $\pm$ 0.36	& $\pm$ 0.46	& $\pm$ 0.55 \\
Large Neptunes & 0.004	& 0.010	& 0.12	& 0.21	& 0.49	& 0.80	& 1.23	& 1.86	& 2.26	& 2.76	& 2.99 \\
& $\pm$ 0.003	& $\pm$ 0.007	& $\pm$ 0.03	& $\pm$ 0.04	& $\pm$ 0.08	& $\pm$ 0.11	& $\pm$ 0.17	& $\pm$ 0.24	& $\pm$ 0.29	& $\pm$ 0.36	& $\pm$ 0.39 \\
Small Neptunes & 0.031	& 0.20	& 0.89	& 2.61	& 5.82	& 10.24	& 15.48	& 19.90	& 23.41	& 26.00	& \nodata \\
& $\pm$ 0.011	& $\pm$ 0.04	& $\pm$ 0.09	& $\pm$ 0.18	& $\pm$ 0.34	& $\pm$ 0.54	& $\pm$ 0.84	& $\pm$ 1.19	& $\pm$ 1.44	& $\pm$ 1.65 & \nodata \\
Superearths & 0.16	& 0.83	& 2.20	& 4.82	& 8.63	& 12.54	& 17.09	& 20.31	& 26.11	& \nodata	& \nodata \\
& $\pm$ 0.03	& $\pm$ 0.12	& $\pm$ 0.23	& $\pm$ 0.53	& $\pm$ 0.79	& $\pm$ 1.12	& $\pm$ 1.61	& $\pm$ 1.95	& $\pm$ 2.87	& \nodata & \nodata \\	
Earths & 0.18	& 0.75	& 2.39	& 4.86	& 7.30	& 9.83	& 13.49	& 16.55 & \nodata & \nodata & \nodata \\			
& $\pm$ 0.09	& $\pm$ 0.17	& $\pm$ 0.44	& $\pm$ 0.69	& $\pm$ 1.04	& $\pm$ 1.40	& $\pm$ 2.27	& $\pm$ 3.60 & \nodata & \nodata & \nodata \\ [+1.5ex]
Any planet & 0.35 & 1.43 & 4.58 & 9.87 & 19.24 & 29.57 & 40.45 & 52.26 & \nodata & \nodata & \nodata \\             
& $\pm$ 0.05  & $\pm$ 0.15   & $\pm$ 0.39  & $\pm$ 0.70  & $\pm$ 1.21 & $\pm$ 1.78 & $\pm$ 2.73 & $\pm$ 4.16 & \nodata & \nodata & \nodata \\ [-2.5ex]
\enddata
%
\tablecomments{The top line in each group represents the number of
stars that have at least one planet in that period range (in percent),
and the bottom line corresponds to the uncertainty.
\label{tab:stars_with_planets}}
%
\end{deluxetable}

%
%


\end{thebibliography}
\end{document}